\documentclass[aps,prl,showpacs,preprintnumbers,amsmath,amssymb,superscriptaddress,twocolumn,10pt]{revtex4-2} %
\usepackage{amsmath}
\usepackage{graphicx}
\usepackage{float}
\usepackage{ulem}
\usepackage{physics}
\usepackage{xr}
\usepackage{CJKutf8}
\usepackage{orcidlink}
\usepackage{comment}
\usepackage{appendix}
\usepackage{caption}
\usepackage{subcaption} %
\usepackage[percent]{overpic}
\newcommand{\panellabelfont}{\footnotesize\bfseries}
\newcommand{\panel}[3]{%
  \refstepcounter{subfigure}\label{#3}%
  \put(#1,#2){{\panellabelfont(\alph{subfigure})}}%
}

\newcommand{\bs}{\boldsymbol}
\newcommand{\ansatz}{(MP)$^2$NQS}

\graphicspath{{./figures/}}

\definecolor{gold}{rgb}{0.83, 0.69, 0.22}

\newcommand{\paul}[1]{\textcolor{gold}{\textbf{YY: #1}}}
\newcommand{\cs}[1]{\textcolor{blue}{\textbf{CS: #1}}}

\newcommand\REMARKS[1]   {\textbf{\textcolor{red}{[#1]}}}
\newcommand\COMMENTED[1] {}

\newcommand{\mr}[1]{\mathrm{#1}}
\def\cA{{\mathcal A}}
\def\cF{{\mathcal F}}

\date{}
\makeatletter
\def\Dated@name{} 
\makeatother
\begin{document}
\begin{CJK*}{UTF8}{gbsn}
\title{Neural Network Discovery of Paired Wigner Crystals in Artificial Graphene}

\begin{abstract}
Moir\'e systems have emerged as an exciting new tunable platform for engineering and probing quantum matter. A large number of exotic states have been observed, 
stimulating
intense efforts in 
experiment, theory, and simulation.
Utilizing a neural-network-based quantum Monte Carlo approach, we discover
a new ground state 
of the two-dimensional electron gas in a honeycomb moir\'e potential
at a filling factor of $\nu_m =1/4$
(one electron every four moir\'e minima).
In this %
state, two opposite-spin electrons pair to form singlet-like valence bond state
which restore local $C_6$ symmetry in hexagonal molecules each spanning $6$ moir\'e minima. These molecules of 
pairs then form a molecular Wigner crystal, leaving one quarter of the  moir\'e minima mostly depleted. 
The formation of such a paired Wigner crystal, absent any confining potential or attractive interaction 
to facilitate  
``pre-assembling' the molecule, provides 
a fascinating case of collective phenomena in strongly interacting quantum many-body systems,
and opportunities to engineer exotic properties.
\end{abstract}
\author{Conor Smith \orcidlink{0009-0005-4580-762X}}
\affiliation{Center for Computational Quantum Physics, Flatiron Institute, New York, NY, 10010, USA}
\affiliation{Department of Electrical and Computer Engineering, University of New Mexico, Albuquerque, NM 87131, USA}
\author{Yubo Yang (杨煜波) \orcidlink{0000-0002-8800-9426}}
\affiliation{Center for Computational Quantum Physics, Flatiron Institute, New York, NY, 10010, USA}
\affiliation{Department of Physics and Astronomy, Hofstra University, Hempstead, NY, 11549, USA}
\author{Zhou-Quan Wan \orcidlink{0009-0007-6260-8715}}
\affiliation{Center for Computational Quantum Physics, Flatiron Institute, New York, NY, 10010, USA}
\author{Yixiao Chen (陈一潇) \orcidlink{0000-0001-8201-5887}}
\affiliation{ByteDance Seed}
\author{Miguel A. Morales \orcidlink{0000-0002-6389-3067}}
\affiliation{Center for Computational Quantum Physics, Flatiron Institute, New York, NY, 10010, USA}
\author{Shiwei Zhang \orcidlink{0000-0001-9635-170X}}
\affiliation{Center for Computational Quantum Physics, Flatiron Institute, New York, NY, 10010, USA}
\maketitle
\end{CJK*}

\subsection{Introduction}
Moiré heterostructures provide a rich and versatile platform for engineering correlated quantum matter~\cite{Kennes_MoireHeterostructuresCondensedmatter_2021,Mak_SemiconductorMoireMaterials_2022,Adak_TunableMoireMaterials_2024a,Andrei_MarvelsMoireMaterials_2021,Checkelsky_FlatBandsStrange_2024a,Nuckolls_MicroscopicPerspectiveMoire_2024}.
Their tunability has enabled the observation of a %
remarkable range of quantum phenomena, including generalized Wigner crystals~\cite{wang_correlated_2020, regan_mott_2020, Li_Continuous_2021, shabani_deep_2021, nieken_direct_2022,Zhou_Electroinc_2025}, unconventional superconductors~\cite{Xia_SuperconductivityTwistedBilayer_2025,Guo_Superconductivity50degTwisted_2025,Jia_AnomalousSuperconductivityTwisted_2025}, and fractional quantum Hall states~\cite{cai_signatures_2023,wang_fractional_2024}.
Despite intense experimental and theoretical efforts, our understanding of their full many-body landscape remains incomplete, in part because %
even the best computational methods still struggle to capture emergent orders hidden in strongly 
correlated quantum many-body systems.

\begin{figure*}[!htb]
    \centering
    \begin{overpic}[width=0.93\linewidth]{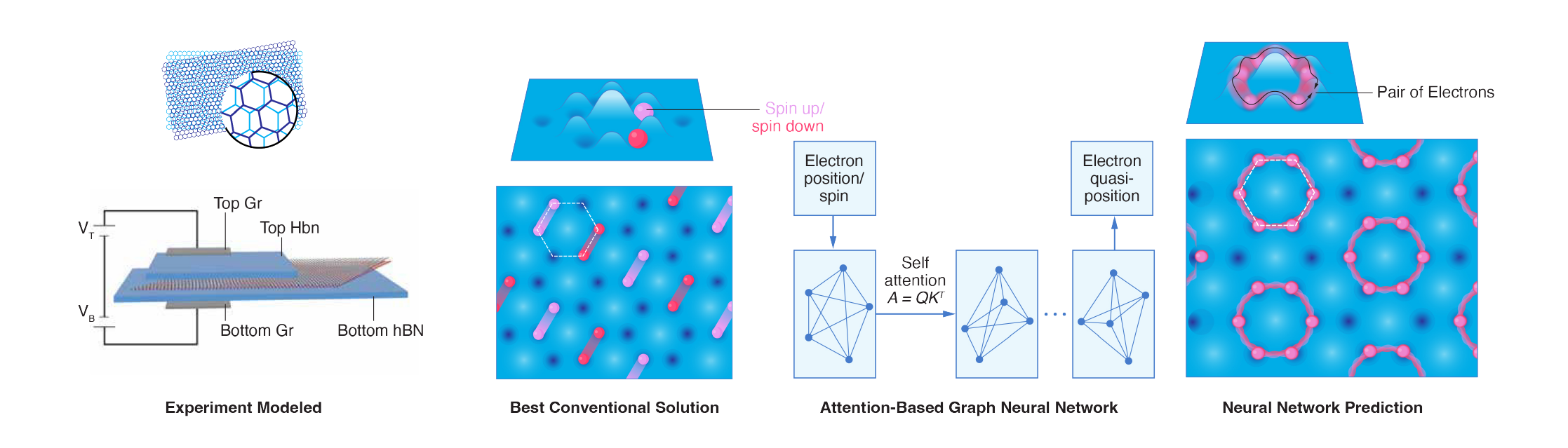}
  \setcounter{subfigure}{0}
  \panel{6}{2}{fig:montage:a}
  \panel{28}{2}{fig:montage:b}
  \panel{48}{2}{fig:montage:c}
   \panel{73}{2}{fig:montage:d}
\end{overpic}

\caption{\textbf{Discovery of the Paired Wigner crystal state.} 
\textbf{a,} Schematics for realization of a honeycomb lattice in a twisted TMD bilayer device.
\textbf{b,} Illustration of the ground state %
obtained using the conventional DMC approach, which has %
each electron distributing over two moir\'e minima and forming an AFM valence bond solid. %
The top panel shows the landscape of the honeycomb moi\'e potential and electron configurations at quarter filling ($\nu_m=1/4$).
\textbf{c,} Illustration of %
our NQS with a backflow neural network, which transforms electron positions into quasi-positions with a self-attention graph neural network.
\textbf{d,} Illustration of the NQS solution, which %
predicts a %
triangular Wigner crystal %
of ring motifs, each made up of a pair of up and down electrons. 
    }
    \label{fig:cartoon}
\end{figure*}

Here, we employ a neural-network quantum Monte Carlo approach~\cite{Pescia_MessagepassingNeuralQuantum_2024,Smith_2024} to explore artificial graphene~\cite{Ma_RelativisticMott_2024}, a honeycomb moir\'e lattice in a two-dimensional semiconductor device, which may be realized as a layer-polarized $\Gamma$-valley transition-metal dichalcogenide (TMD) bilayer~\cite{Angeli2021}.
Our approach incorporates neural networks into a physically motivated variational ansatz, enabling computational discovery 
with a balance of strong predictive power and access to sufficiently large system sizes. 
Through it, we uncover an unexpected ground state: %

electrons of opposite spins bind into singlet %
pairs with their wavefunctions spread across local hexagons, restoring C$_6$ symmetry. These pairs themselves crystallize into a triangular lattice,
forming a genuine molecular crystal.
Different from the known Wigner molecular crystals~\cite{Li_WignerMolecularCrystals_2024,li2024emergentwignerphasesmoire}, this state occurs at a low 
filling ($\nu_m = 1/4$, 
i.e., one electron every four moiré minima),
without any confining potential or attractive interactions to drive 
a tendency for molecular formation. 
An overview of our work and illustration of the physics of the new state are presented in Fig.~\ref{fig:cartoon}.

The state we discover is reminiscent of the paired Wigner crystal (PWC), which was proposed decades ago as a candidate ground state of the homogeneous electron gas
~\cite{Moulopoulos_NewLowDensity_1992,Taut_WignerCrystallizationTwodimensional_2001},
but was %
shown to be unstable %
with more accurate calculations~\cite{Drummond_PhaseDiagramLowDensity_2009}.
The spontaneous formation of this exotic quantum state in a moir\'e setting demonstrates a new route to stabilizing molecular electronic crystals.
It is unexpected, not only because the state is not what conventional state-of-the-art computational methods would predict, but also because the physics is surprising. 
The state combines %
competing and seemingly incompatible elements: kinetic-energy-driven valence bond formation and pairing within each molecule, and %
interaction-driven Wigner crystallization of the molecular centers.
Moreover, the paired state develops %
out of a 
semi-metallic phase and remains non-magnetic even deep in the insulating regime, 
which is contrary to expectations based on similar settings under other parameter settings (filling~\cite{Yang_FerromagneticSemimetalChargeDensity_2024} or moir\'e geometry~\cite{Yang_MetalInsulatorTransitionSemiconductor_2024}) and different from what 
prior state-of-the-art computational methods yield, as we further discuss below.
This finding not only expands the catalog of moir\'e phases but also illustrates the power of neural-network approaches to reveal unexpected quantum many-body phenomena. The PWC’s structure and dynamics suggest rich experimental signatures and open the door to further exotic phases such as paired supersolids.

\COMMENTED{
\paul{I moved the Hamiltonian to the supplemental}
\section{Results}
We study the ground state of holes in a TMD heterobilayer,
a cartoon of which  is shown in Fig.~\ref{fig:cartoon}(a), using the moir\'e continuum Hamiltonian 
\begin{equation}\label{eq:mch}
\hat{H} = -\frac{1}{2}\sum_i \nabla^2_i - V_m/W \sum_i \Lambda(\bs{r}_i) +r_s \sum_{i<j} \frac{1}{|\bs{r}_i-\bs{r}_j|},
\end{equation}
which assumes that the holes are confined to a single layer while the presence of the other layer can be modeled by an effective potential with shape $\Lambda(\bs{r}\,) = 2\sum^3_{j=1} \cos(\bs{r}\cdot \bs{g}_j + \phi)$. %
The vectors $\bs{g}_j$ are defined as the moir\'e unit cell's three smallest reciprocal lattice vectors and $\phi$ controls the shape of the potential, which we set to $\phi=60^\circ \pm n 120^\circ$ to model a honeycomb potential. 
The effective Bohr radius $a_B^* \equiv \frac{\hbar^2}{\vert m^*\vert} / \frac{e^2}{4\pi\epsilon}$ is used as length unit and
the Fermi energy of the unpolarized gas $W=\frac{\hbar^2}{\vert m^*\vert r_s^2}$ is used as the energy unit.
All values are given in Hartree atomic units.
The parameter $V_m$ controls the depth of the moir\'e potential. 
}

\subsection{Phase Discovery with the Neural Quantum State}

\begin{figure*}[ht]
\begin{overpic}[width=\linewidth]{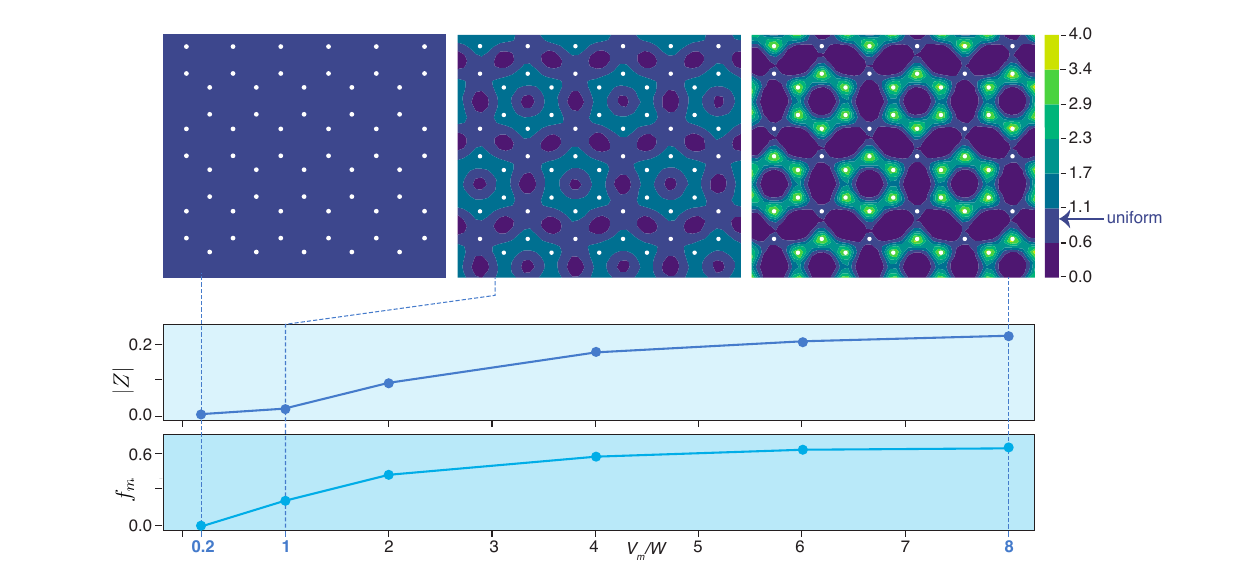}
  \setcounter{subfigure}{0}
  \panel{15.8}{22}{fig:montage:a}
  \panel{39.7}{22}{fig:montage:b}
  \panel{77}{22}{fig:montage:c}
\end{overpic}
\caption{\textbf{Metallicity and %
molecular localization.} \textbf{a,b,c,} The electron densities for three 
values of $V_m/W$, as indicated by the dashed lines.
Metallicity is measured via the complex polarization $\vert Z\vert$, which is zero for a metal and asymptotically unity for 
an insulator. The molecular localization $f_m$ measures the degree of localization on the ring motifs, %
characterized by the ratio of the integrated charge density inside and outside the ring. This ratio will reach unity if no electron density ``leaks out of a molecule''. 
At low moir\'e potential depth, both quantities are zero, giving a metallic state with uniform density and no molecular formation.
As the moir\'e potential deepens, both metallicity and localization %
become finite %
as the paired Wigner crystal forms.}
\label{fig:evo_ring}
\end{figure*}

Figure~\ref{fig:cartoon} summarizes the experimental context and the central 
result of this work.
Figure~\ref{fig:cartoon}(b) shows the best ground state obtained using the existing state-of-the-art approach: Diffusion Monte Carlo (DMC) projection using an optimized Slater–Jastrow–backflow wavefunction 
based on orbitals derived from density-functional theory (DFT) with a hybrid functional under
the local-density approximation (LDA) the details of which are in the SM.
In this ``conventional'' solution, each electron partially occupies two neighboring moir\'e minima, while opposite-spin electrons localize on far ends of a hexagonal plaquette.
This configuration breaks $C_6$ rotation symmetry and realizes a valence-bond solid state of an antiferromagnetic insulator. In contrast, the true ground state,
revealed by the NQS solution, 
is qualitatively different,
as illustrated in Fig.~\ref{fig:cartoon}(d). %
Here, opposite-spin electrons bind into singlet pairs whose wavefunctions delocalize coherently over an entire hexagonal ring of six moir\'e minima, reminiscent of a resonating valence bond state. These singlet molecules, which fully restore $C_6$ symmetry, %
then %
crystallize into a triangular lattice, forming a paired Wigner crystal (PWC). 

The discovery of this novel state not only 
underscores the power of NQS but also provides important guidance for effectively harnessing its promise. 
The PWC state, 
which is entirely missed by conventional DMC starting from %
mean-field orbitals, 
requires the flexible and expressive
wavefunction representations provided by NQS. 
In order to confirm that the state is not an artifact of the variational ansatz, we performed a sequence of tests combining NQS with DMC, as
further discussed in the SM.
In brief, we use the final NQS orbitals to construct a Slater determinant,
and feed the trial wavefunction derived from it into an independent DMC calculation. The PWC state remains robust in the new DMC, which %
yields a {\it lower\/} energy than the state obtained from the ``conventional'' DMC calculations pictured above in Fig.~\ref{fig:cartoon}(b).
Since DMC is variational, 
this indicates %
that the PWC %
state represents a genuine improvement of the nodal surface
of the ground-state wave function. %
As NQS studies experience rapid growth, validation tests such as these will have an important role in ensuring solutions are not merely artifacts of the neural architecture and to provide increased confidence in the predictive power and relevance of this approach to experimentalists and theorists alike.

\subsection{Properties}

\begin{figure}[ht]

\begin{overpic}[width=\linewidth]{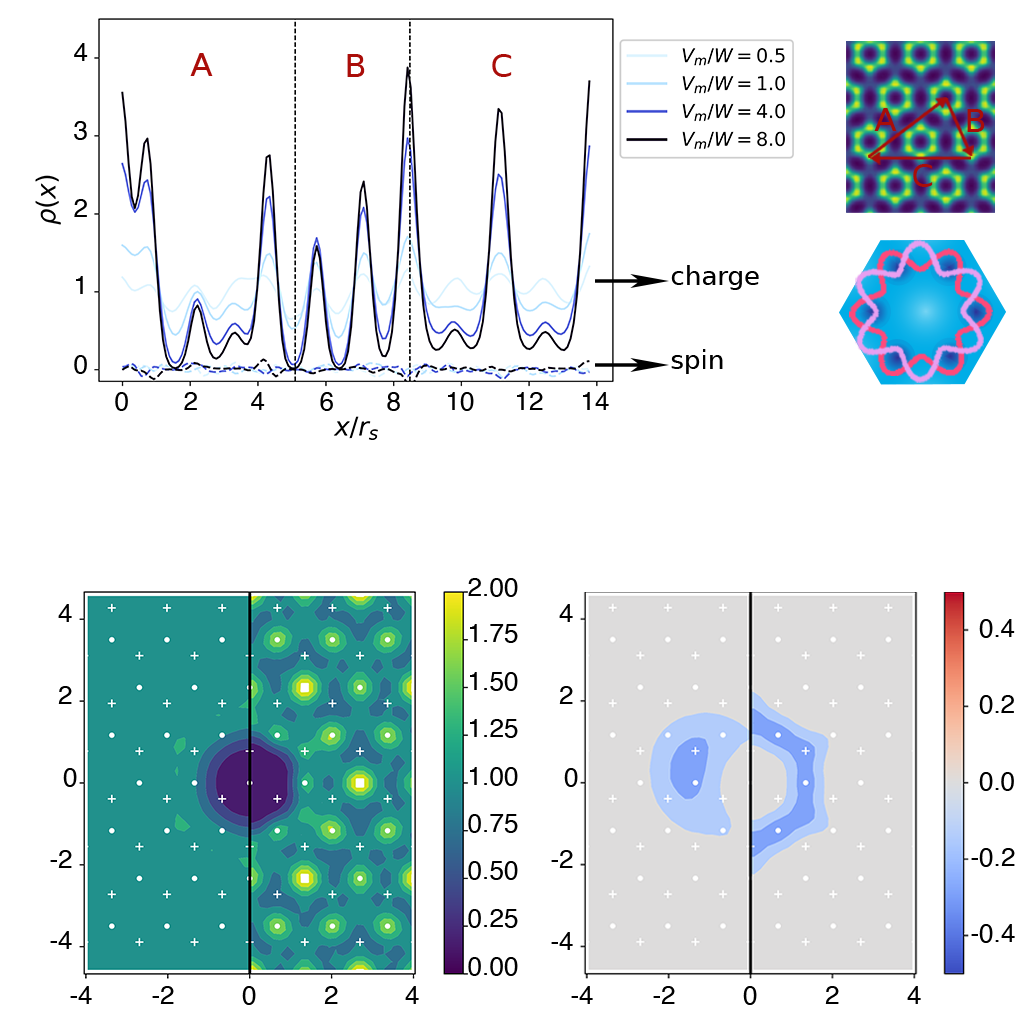}
  \setcounter{subfigure}{0}
  \panel{0}{55}{fig:montage:a}
  \panel{0}{0}{fig:montage:b}
  \panel{50}{0}{fig:montage:c}
\end{overpic}

\caption{\textbf{Charge and spin densities and correlation functions.}
\textbf{a,} The 
charge (solid) and spin (dashed) densities are plotted 
along the path depicted in the top-right inset for various potential depths. The color indicates the value of $V_m/W$ as shown in the legend box. The average charge and spin densities are indicated by the arrows. 
The bottom right inset shows A schematic illustration of the singlet spin state of each molecule at large $V_m/W$. 
\textbf{b,} Charge-charge correlation function with $V_m/W=0.5$ on the left half and $V_m/W=8.0$ on the right half;
\textbf{c,} the corresponding spin-spin correlations.
}
\label{fig:cs-correlation}
\end{figure}

We systematically investigate the properties of the ground state as the moir\'e potential depth is varied. Through a number of different probes, we establish that
there are at least two different phases as a function of $V_m/W$. 
At low  $V_m/W$,
the system preserves lattice translation symmetry. 
As $V_m/W$ is increased, this symmetry is broken, and at sufficiently large values a molecular Wigner crystal
develops, which is the state described above. 
In Fig.~\ref{fig:evo_ring}, panels (a-c) show the evolution of the charge density,
from a nearly uniform liquid at $V_m/W=0.2$, to weak molecular modulations at $V_m/W=1$, and to well-formed hexagonal molecules at $V_m/W=8$.
This is illustrated further in Fig.~\ref{fig:cs-correlation}(a) which shows linecuts of the charge density for a sequence of  $V_m/W$ values. 

Below, we further 
quantify these states from several perspectives. 
The system is metallic at low $V_m/W$ and the molecular Wigner crystal at large $V_m/W$ is a PWC state.
In the latter, the molecules are in a singlet,
as illustrated in Fig.~\ref{fig:cs-correlation}(c). 
Indeed, the local spin remains $S_z({\mathbf r})=0$ throughout the system regardless of the moir\'e potential depth, as shown in
Fig.~\ref{fig:cs-correlation}(a). We examine electron-electron charge and magnetic correlations. Their short-range behavior illuminates how the exchange-correlation hole 
of the 2DEG is influenced by the moir\'e potential; the long-range behavior further
establishes the liquid versus crystal states. We probe the internal structure of 
the molecule and the intermolecular correlations in the PWC phase, which exhibit a fascinating 
interplay between the electron pairing and long-range order as the moir\'e potential depth is tuned.

\subsubsection{Metallicity and Molecular Localization} 

\COMMENTED{
The PWC develops continuously as the moir\'e potential depth $V_m$ is increased. Figure~\ref{fig:evo_ring}(a-c) show the evolution of the charge density from a nearly uniform liquid at $V_m/W=0.2$, to weak molecular modulations at $V_m/W=1$, and to well-formed hexagonal molecules at $V_m/W=8$. \paul{mention linecuts here}
}

To quantify metallicity, we calculate the complex polarization~\cite{Resta_ElectronLocalizationInsulating_1999, Souza_PolarizationLocalizationInsulators_2000}
\begin{equation} \label{eq:Z_cpol}
Z = \langle\psi|\mathrm{exp}\left (-i \sum^{N_e}_{i=1} \bs{g}\cdot \bs{r}_i \right )  |\psi \rangle,
\end{equation}
where $\bs{g}$ is the smallest reciprocal-space lattice vector commensurate with the simulation cell. In the thermodynamic limit, this quantity is 0 in a metal and $\vert Z\vert\rightarrow 1$ in an insulator.
The $\vert Z\vert$ values here, which are computed in a finite simulation cell, are affected by finite-size effects and range between 0 and 1.
Despite the onset of visible molecular structure at $V_m/W=1$, $\vert Z\vert$ remains close to zero, indicating that the system remains largely delocalized. Only at $V_m/W>1$ does $\vert Z\vert$ increase significantly, which signals a metal-insulator transition (MIT).

To quantify the degree of localization associated with molecular formation, we partition space according to those 
moir\'e minimum sites
belonging to hexagonal rings with a molecule (``occupied'') and those outside (``unoccupied''). The occupancy of each moir\'e minimum site is defined as the integrated charge density of the Voronoi cell surrounding it
\begin{align}
f_o =& \frac{1}{N_o} \sum_{i\in \text{occupied}} \rho(\mathcal{V}_i) \\
f_u =& \frac{1}{N_u}\sum_{i\in \text{unoccupied}} \rho(\mathcal{V}_i).
\end{align}
At filling $\nu_m=1/4$, perfect molecule formation, which means all the electron densities are contained in the occupied rings with no ``leakage'' to the outside sites, corresponds to $f_u=0$ and $f_o=1/3$, whereas a uniform charge distribution yields $f_u=f_o=1/4$. We thus define the molecular localization measure
\begin{equation}
f_m = 3(f_o-f_u),
\end{equation}
which ranges from $0$ for a uniform state to $1$ for perfectly localized molecules. 
As seen %
in Fig.~\ref{fig:evo_ring}, $f_m$ begins 
to grow already at $V_m/W=1$, seemingly %
before the MIT suggested %
by $\vert Z\vert$. The value reaches $\sim 0.6$ at $V_m/W=8$, indicating that there is still substantial ``leakage'' of the molecule outside the ring.

\COMMENTED{

\REMARKS{ move this somewhere further below? }
These observations establish at least two different phases as the potential depth is tuned. 
Pair formation and symmetry breaking occur within a semi-metallic state, and the PWC with global charge localization follows at larger moir\'e potential strength. 
In between it is intriguing to wonder about superconductivity or
the tantalizing possibility of supersolid-like behavior.
A more systematic study, with sharper detection measuers and in larger system sizes, is required to address these conclusively. We have used projected-BCS type of wave fucntions to conduct a preliminary exploration of this question. 
\REMARKS{complete - refer to SM}

\REMARKS{weakened these statements, since size effect is very subtle}
Together, these observations establish a two-stage evolution of the PWC: (i) pair formation and symmetry breaking occur within a semi-metallic state, and (ii) global charge localization follows at larger moir\'e potential strength. It is important to note that no such continuous MIT is predicted in conventional theories. LDA favors a ferromagnetic metal regardless of moir\'e depth, whereas hybrid LDA predicts an insulator even in absence of the moir\'e potential, and DMC[LDA] gives an abrupt transition between the liquid and the AFM insulator (see details in supplemental materials). None of these approaches capture the gradual emergence of paired molecular order revealed by the NQS.
}

\subsubsection{Charge and Spin Correlation} 

Figure~\ref{fig:cs-correlation}(b) %
contrasts the liquid and PWC phases 
by the charge %
at two representative values of the 
moir\'e potential depth. To do so, we use the charge pair correlation function, $g(\bs{r})$, defined as
\begin{align} \label{eq:g} 
g(\bs{r}) =  \dfrac{1}{A} \int d\bs{r}' \sum_{\sigma,\sigma'} g_{\sigma,\sigma'}(\bs{r}, \bs{r}+\bs{r}'),
\end{align}
where the spin- and pair-resolved contributions are defined as
\begin{align}
g_{\sigma_1,\sigma_2}(\bs{r}_1, \bs{r}_2) &= \frac{\Big\langle 
\sum_{i\neq j} \delta_{\sigma_1,\sigma_i} \delta(\bs{r}_1-\bs{r}_i)\,  \delta_{\sigma_2,\sigma_j} \delta(\bs{r}_2-\bs{r}_j)\Big\rangle}{\rho_{\sigma_1}(\bs{r}_1)\,\rho_{\sigma_2}(\bs{r}_2)}, 
\end{align}
with $\rho_{\sigma}(\bs{r})=\langle \sum_i \delta_{\sigma,\sigma_i} \delta(\bs{r}-\bs{r}_i) \rangle$.
At $V_m/W=0.5$,
the behavior of $g({\mathbf r})$
is that of a liquid, with a short-range exchange-correlation hole slightly 
modified by the weak moir\'e potential modulation.
This is consistent with the charge densities of 
Fig.~\ref{fig:evo_ring}(a) and Fig.~\ref{fig:cs-correlation}(a)  
discussed earlier.
At $V_m/W=8.0$, the charge–charge correlation function 
exhibits long-range order consistent with a triangular lattice of molecular centers in a PWC state.

Spin correlations are studied in Fig.~\ref{fig:cs-correlation}(c)
for the same two systems using the spin pair correlation function
\begin{align}
g_s(\mathbf r) = \dfrac{1}{A} \int d\bs{r}' \sum_{\sigma,\sigma'}(2\delta_{\sigma,\sigma'} - 1)  g_{\sigma,\sigma'}(\bs{r}, \bs{r}+\bs{r}').
\end{align}
As shown in Fig.~\ref{fig:cs-correlation}(a),  there is no net spin density anywhere.
However, even in the liquid phase,
the spin correlation shows a pronounced short-range pattern that breaks C$_6$ symmetry. 
This reflects the nematic spin correlation recently seen~\cite{Smith_2024} in the 2DEG at intermediate density between the Wigner crystal and the high-density Fermi liquid, when the system is a correlated liquid.  
In the PWC phase,
the spin–spin correlation function %
remains short-ranged, with correlations confined to distances comparable to the diameter of a single hexagonal molecule. 
We further examine the intra- and inter-molecule spin correlations next.

\COMMENTED{
The coexistence of long-range charge order and short-range spin correlations is another %
key signature of the PWC, %
distinguishing it from a conventional Wigner crystal or an antiferromagnetic valence bond solid.
}

\COMMENTED{
Despite the strong charge localization in the PWC regime, the system does not develop magnetic order. As shown in Fig.~\ref{fig:cs-correlation}(a),  the spin density remains at zero even as strong charge density modulation is established in deep moir\'e potentials. \paul{This suggests singlet pairing, which we verify by examining correlation functions.}%
The charge–charge correlation function exhibits long-range order consistent with a triangular lattice of molecular centers (Fig.~\ref{fig:cs-correlation}(c)), while the spin–spin correlation function remains short-ranged, with correlations confined to distances comparable to the diameter of a single hexagonal molecule. The coexistence of long-range charge order and short-range spin correlations is a defining signature of a singlet molecular crystal, distinguishing it from a conventional Wigner crystal or an antiferromagnet.
\paul{mention connection to the nematic spin-correlated liquid.}
}

\begin{figure*}[!htbp]
\centering
\begin{overpic}[width=0.93\linewidth]{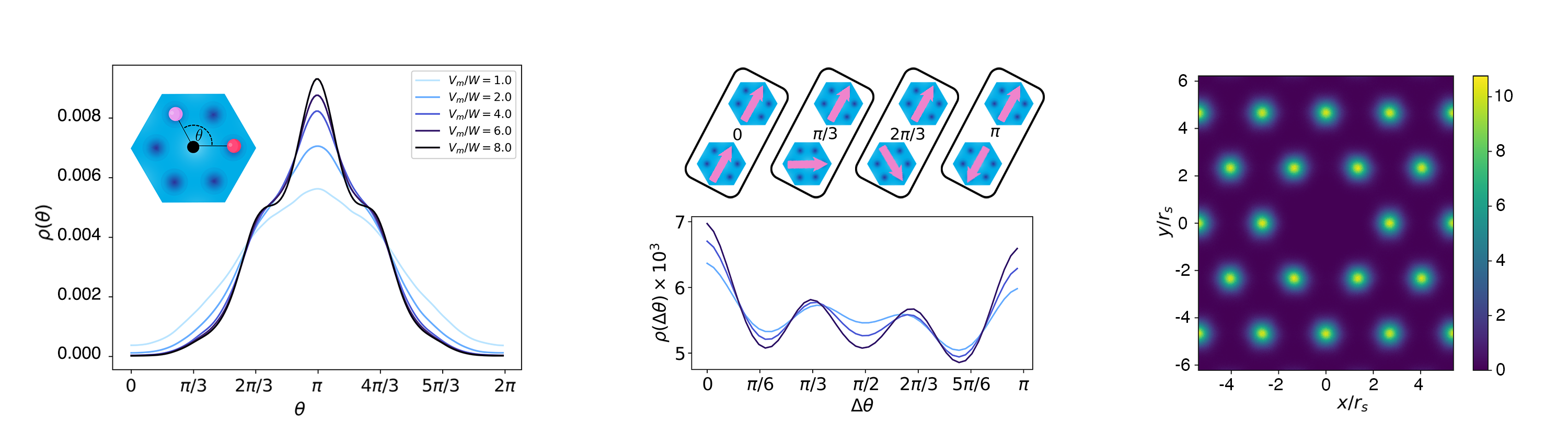}
  \setcounter{subfigure}{0}
  \panel{3}{2}{fig:montage:a}
  \panel{38}{2}{fig:montage:b}
  \panel{70}{2}{fig:montage:c}
\end{overpic}
\caption{\textbf{Structure and correlations of the pairs.}
\textbf{a,} Intra-molecule correlation between the
pair of up and down electrons on the same ring is characterized by 
the %
distribution of their relative angle $\theta$, as illustrated by the inset.
Curves of different colors show different moir\'e  well depths.
\textbf{b,} Orientational correlation between the nearest-neighbor pairs. %
The colors of the curves indicate potential well depths, as given in the legend in panel \textbf{a.} 
The configuration corresponding to each peak 
is depicted at the top of the plot.
\textbf{c,} Correlation between the center-of-mass positions.
}
\label{fig:pair_obs}
\end{figure*}

\subsubsection{Molecular Structure and Correlations of the PWC} %

To probe the internal structure of the singlet pairs, we first measure %
the angular separation between the two electrons in the same molecule. 
As we have seen from the charge densities, C$_6$ symmetry is preserved, so their 
angular structure can be reduced to the relative angle, as depicted in the inset 
in Fig.~\ref{fig:pair_obs}a. The results are shown in the main plot for 
a series of  moir\'e potential depths. We see the
two opposite-spin electrons tend to stay on the far ends of the same molecule ($\theta=\pi$), %
maximizing their separation. 
Configurations at nearby angular separations ($\theta=2\pi/3, 4\pi/3$) occur with much reduced probability, while configurations in which both electrons occupy the same side of the ring ($\theta=\pi/3, 5\pi/6$) are strongly suppressed despite the presence of moir\'e minima there.  
These observations motivate a natural picture %
of each singlet molecule
as an effective electric dipole, characterized by a center-of-mass (COM) position $\bs{c}_i$ and a dipole moment $\bs{p}_i$.
The dipoles are not rigidly locked to minima and can ``rotate'' freely to restore C$_6$ symmetry. 
As the potential depth is increased, the dipole moment increases and the structure of the dipole becomes sharper, as seen by the rise of the central peak with $V_m/W$.

The COMs of the molecules form a triangular lattice, as seen directly in their pair correlation function, $g(\bs{c}_i-\bs{c}_j)$, shown in Fig.~\ref{fig:pair_obs}c. To determine if the dipole moments order, we %
examine the alignment angle between neighboring singlets by computing $\cos(\Delta\theta)=\hat{\bs{p}}_i\cdot\hat{\bs{p}}_j$. Perfect alignment and anti-alignment of the pairs are characterized by 
$\Delta\theta=0$ and $\pi$, respectively.
Figure~\ref{fig:pair_obs}b shows the distribution of alignment angles for nearest neighbor  %
singlets at several moir\'e depths.

\subsubsection{Potential Other Phases} 
These observations establish at least two different phases as the potential depth is tuned. Starting from the metallic liquid state at $V_m/W=0$,
pair formation and symmetry breaking occur with increasing $V_m$ leading to a PWC with global charge localization at large moir\'e potential strength. 
In between, it is intriguing to wonder about superconductivity and
the tantalizing possibility of supersolid-like behavior. This is suggested by the observation that, at $V_m/W=1$, we see signatures of pair formation while the complex polarization remains consistent with a metallic state.
We have used projected-BCS type of wave functions to conduct a preliminary exploration of this question. In our NQS-backflow ansatz, we have 
mostly used a determinant structure~\cite{Smith_2024} but when the Slater determinant 
is replaced by a projected-BCS structure, the energy is further lowered. At intermediate $V_m/W\approx 4$, the largest lowering is observed as shown in Table~\ref{tab:polar_moire}.
A more systematic study, with sharper detection measures and in larger system sizes, is required to address these conclusively.

\subsection{Conclusion and Outlook}
Using an expressive NQS tailored to perform unbiased phase discovery in strongly correlated systems, we discover %
a paired molecular Wigner crystal phase in a honeycomb moir\'e potential at $\nu_m=1/4$. In this exotic state, opposite spin electrons form strongly correlated pairs whose charge densities delocalize over 
multiple moir\'e minima, while the resulting molecular centers crystallize into a triangular lattice. This intertwines short-range valence-bond pairing and long-range Wigner crystallization in  a molecular electron solid.
The state %
emerges in a system with purely repulsive Coulomb interaction 
and without any direct 
confinement for molecular formation.
As such it is distinct from  the rich set of states that have been seen in strongly correlated electrons.

This system is well within the realm of current experimental setups in 2D materials. 
Over a wide range of potential depths, 
we find long-range charge order together with short-range spin correlations in the form of singlet pairing. This prevalence and robustness will inspire theoretical exploration~\cite{Zverevich2026} and facilitate experimental detection.
Given the rapid advances in experiment, including the realization of artificial graphene~\cite{Ma_RelativisticMott_2024}, probing of magnetism~\cite{Tang_EvidenceFrustratedMagnetic_2023}, and real-space imaging advances~\cite{Tsui_DirectObservationMagneticfieldinduced_2024,Li_WignerMolecularCrystals_2024,Xiang_ImagingQuantumMelting_2025,Ge_VisualizingImpactQuenched_2025,Berger_ImagingElectronHoleAsymmetry_2025} it should now be possible to search for and detect signatures of the PWC in the charge and magnetic textures.

Our results expand the landscape of correlated phases in moir\'e materials and suggest %
tantalizing new possibilities. 
More broadly, our findings highlight the importance of expressive and systematically improvable methods for phase discovery.
The PWC state is a surprising discovery
as it is different from what state-of-the-art computation 
would have predicted without the neural network enhancement.
This motivates the further development of NQS and other many-body computation methods, and their 
application for predictive computations in not only  moir\'e but other quantum platforms and materials.

\section{Acknowledgments}
The Flatiron Institute is a division of the Simons Foundation.
We thank Ilya Esterlis for useful discussions.
YY acknowledges support by the National Science Foundation (NSF) under grant number DMR-2532734 and the use of the Hofstra Star HPC cluster, acquired with support from the NSF Major Research Instrumentation (MRI) program under grant number 2320735.
\clearpage
\onecolumngrid
\section*{Supplementary Information}

\setcounter{section}{0}
\renewcommand{\thesection}{S\arabic{section}}
\setcounter{figure}{0}
\renewcommand{\thefigure}{S\arabic{figure}}
\setcounter{table}{0}
\renewcommand{\thetable}{S\arabic{table}}

\section{Hamiltonian}

We study the ground state of holes in a semiconductor heterobilayer,
a cartoon of which is shown in Fig.~1a of the main text, using the following moir\'e continuum Hamiltonian 
\begin{equation}\label{eq:mch}
\hat{H} = -\frac{1}{2}\sum_i \nabla^2_i - V_m/W \sum_i \Lambda(\bs{r}_i) +r_s \sum_{i<j} \frac{1}{|\bs{r}_i-\bs{r}_j|},
\end{equation}
which assumes that the holes are confined to a single layer while the presence of the other layer can be modeled by an effective potential of the form $\Lambda(\bs{r}\,) = 2\sum^3_{j=1} \cos(\bs{r}\cdot \bs{g}_j + \phi)$. %
The vectors $\bs{g}_j$ are defined as the moir\'e unit cell's three smallest reciprocal lattice vectors and $\phi$ controls the shape of the potential, which we set to $\phi=60^\circ \pm n 120^\circ$ to model the honeycomb potential realized in artificial graphene.
The parameter $V_m$ controls the depth of the moir\'e potential.
The effective Bohr radius $a_B^* \equiv \frac{\hbar^2}{\vert m^*\vert} / \frac{e^2}{4\pi\epsilon}$ is used as length unit and
the Fermi energy of the unpolarized gas $W=\frac{\hbar^2}{\vert m^*\vert r_s^2}$ is used as the energy unit.
All values are given in Hartree atomic units.
The Wigner-Seitz radius in Bohr units, $r_s$, controls the strength of Coulomb interaction relative to the kinetic energy.
 
\section{The Ansatz}
\label{sec:Ansatz}
We largely use an identical ansatz as described in Ref.~\cite{Smith_2024}. The main idea is to construct a determinant of orbitals built with a learnable linear combination of planewaves:
\begin{equation}
    \phi_a(\bs{q}_b) = \sum^{N_k}_{k=1} c_{ak}\exp(i\bs{G}_k\cdot \bs{q}_b)
\end{equation}
where $c_{ak}$ is learnable and $\bs{q}_b$ is a quasiposition generated by a backflow transformation as:
\begin{equation}
    \bs{q}_b = \bs{r}_b + \mathcal{N}(\bs{R})
\end{equation}
with $\bs{r}_b$ denoting the position of particle $b$ and $\bs{R}$ denoting all particle positions. 

The backflow transformation is based on the message passing neural network developed in Ref.~\cite{Pescia_MessagepassingNeuralQuantum_2024}. The idea is to iteratively update one-body and two-body hidden vectors, $\bs{h}_i$ and $\bs{h}_{ij}$, with an attention mechanism. In more detail, the state at layer $t$ is given by

\begin{align}
\bs{g}^{(t)}_i
  &= \Big[\bs{h}^{(t-1)}_i\Big] \,,
&\qquad
\bs{g}^{(t)}_{ij}
  &= \Big[\bs{v}_{ij},\ \bs{h}^{(t-1)}_{ij}\Big] \,\\
\bs{h}^{(t)}_i
  &= \cF_{1}^{(t)}\Big( \sum_{j \neq i} \bs{m}^{(t)}_{ij},\ \bs{g}^{(t)}_i \Big)
     + \bs{h}^{(t-1)}_i \,,
&\qquad
\bs{h}^{(t)}_{ij}
  &= \cF_{2}^{(t)}\Big( \bs{m}^{(t)}_{ij},\ \bs{g}^{(t)}_{ij} \Big)
     + \bs{h}^{(t-1)}_{ij} \,\\
\quad \bs{m}^{(t)}_{ij}
  &= \cA_{ij}^{(t)}\Bigl( \bs{g}_{ij}^{(t)} \Bigr)\odot \cF_{m}^{(t)}\Bigl( \bs{g}_{ij}^{(t)} \Bigr) \,,
&&
\end{align}
where $\cF_{h_1}^{(t)}$, $\cF_{h_2}^{(t)}$ and $\cF_{m}^{(t)}$ are multi-layer perceptrons (MLPs). The visible state $\bs{v}_{ij}$ is a periodized displacement and distance vector given by,

\begin{equation}
    \bs{v}_{ij} = \Big[ \cos(2\pi A^{-1} \bs{r}_{ij}),\ \sin(2\pi A^{-1} \bs{r}_{ij}),\ ||\sin(\pi A^{-1} \bs{r}_{ij})||,||\cos(\pi A^{-1} \bs{r}_{ij})||,s_{ij} \Big]
\end{equation}
with $A$ being the lattice vectors of the supercell and $s_{ij}= 2\delta_{\sigma_i\sigma_j-1}$ where $\sigma_i$ is the spin of particle $i$. The message, $m_{ij}$, is the element-wise multiplication of an MLP transformation of $g_{ij}$ with an attention matrix defined as
\begin{alignat}{2}
\cA_{ij}^{(t)}
  &= \mr{Linear}^{(t)} \circ \mr{GELU}\pqty{\frac{1}{\sqrt{N}} \sum_{l}^{N} \bs{q}_{il}^{(t)} \bs{k}_{lj}^{(t)}} \, ,\qquad &&\\
\bs{q}_{ij}^{(t)}
  &= W_q^{(t)} \cdot \bs{g}_{ij}^{(t)} \, ,
&\hspace{1.2em}\bs{k}_{ij}^{(t)}
  &= W_k^{(t)} \cdot \bs{g}_{ij}^{(t)} \, .
\end{alignat}
To describe the backflow, we simply take the final one-body hidden state, $\bs{h}^{(L)}_i$, with $L$ denoting the last layer, and apply one more MLP so that:
\begin{equation}
    \mathcal{N}(\vec{R}) = \cF_{bf}(h^{(L)}_i).
\end{equation}

The Jastrow is also derived from the same one-body hidden state via another MLP:

\begin{equation}
    \mathcal{J}_\theta(\vec{R})=\sum_i \cF_{J}([\bs{v}_i,h^{(L)}_i]).
\end{equation}
We explicitly break translational invariance in the Jastrow by including one-body positional information, $\bs{v}_i=[\sin(2\pi A^{-1}\bs{r}_i),\cos(2\pi A^{-1}\bs{r}_i)]$ with $A$ being the matrix of lattice vectors of the supercell. This choice is motivated by the fact that the moir\'e potential breaks continuous translational invariance (while remaining periodic).

We can compactly write the full wavefunction as
\begin{equation}
\Psi(\vec{r\,}) =
\text{det}  \left[ 
\{\langle \bs{q}_j; \sigma_j|
 \phi_a; \chi_a\rangle \}\right]
\exp\big( \mathcal{J}_{cck}(\vec{R})+\mathcal{J}_\theta(\vec{R}) \big)\,,
\label{eq:ansatz-0}
\end{equation}
where orbital $\chi_a$ is an eigenstate of $\hat{s}_z$ allowing us to write the wavefunction as product of spin-$\uparrow$ and spin-$\downarrow$ determinants. Finally, $\mathcal{J}_{cck}$ is the analytical form defined by Ceperley, Chester, and Kalos~\cite{CCK1976}.

\begin{table}[]
\centering
\begin{tabular}{|l|l|l|}
\hline
              & dimensions & layers \\ \hline
Iterations &            & 3      \\ \hline
$W_{k/q}$    & 32         & 1      \\ \hline
$\cF_m$    & 32         & 1      \\ \hline
$h_i/\cF_1$          & 32         &  2      \\ \hline
$h_{ij}/\cF_1$     & 26         &  2      \\ \hline
$\mathcal{J}$ & 32         & 3      \\ \hline
\end{tabular}
\caption{Neural Network Hyperparameters}
\end{table}

\section{Optimization and Sampling}
We use the SPRING algorithm \cite{Goldshlager_2024} to update the parameters as:
\begin{equation}
    d\theta_t = (S+\lambda I)^{-1}(g+ \lambda \mu d\theta_{t-1})
\end{equation}
\begin{equation}
    \theta_{t+1}= \theta_t - \eta d\theta_t
\end{equation}
where $\theta_t$ are the parameters of the NQS at step $t$, $S$ is the quantum Fisher matrix and $g$ are the gradients. The learning rate is given by $\eta$, $\lambda$ is the damping term applied to the Fisher matrix, and $\mu>0$ introduces a term analogous to momentum in first-order methods while $\mu=0$ reduces to standard stochastic reconfiguration~\cite{Sorella_1998}. To invert the S matrix we use minSR~\cite{chen2023efficientnumericalalgorithmlargescale,Chen_2024}. Hyperparameters are given in Table~\ref{tab:SR}.

\begin{table}[]
\centering
\begin{tabular}{|l|l|l|}
\hline
$\lambda$     & $1\times10^{-3}$          \\ \hline
$\mu$        & $0.9$                 \\ \hline
$\eta$        & $O(0.1)$                 \\ \hline
Decay         & $1000$    \\ \hline
Samples         & $1024$    \\ \hline
\end{tabular}
\caption{Optimization Hyperparameters}
\label{tab:SR}
\end{table}

To sample from the wavefunction we use Metropolis Adjusted Langevin Algorithm~\cite{besag1994comments} where samples are proposed according to:
\begin{equation}
    \tilde{r} = r + \tau\nabla\log\vert \psi(r) \vert^2 + \sqrt{\tau}\varepsilon
\end{equation}
 and $\varepsilon$ is randomly drawn from a Gaussian distribution. The step size, $\tau$, is tuned adaptively to target a harmonic average of acceptance rate at $65\%$. We propose and accept/reject 20 updates and use only the last to calculate energies and gradients when training the wavefunction.
\section{BCS-Type Wavefunction Details}\label{sec:moire_NNDetails}
In this section, we present the BCS-type wavefunction referenced in the main text. We stress here that this wavefunction is not necessary to find the PWC but it does produce a lower energy than the determinant based ansatz. This is important because it is not that our work is biased to encourage pairing but rather that the apparent pairing motivated us to implement a wavefunction which can describe pairing more explicitly. To explain the form of this wavefunction, we begin with an unpaired determinant of two spin species which we can express as a product of two determinants:

\begin{equation}
\label{eqn:spin_split}
    \Psi(\vec{x}) = \left | \Phi^{(\uparrow)}\left [\vec{x}^{(\uparrow)}\right ] \right | \cdot \left | \Phi^{(\downarrow)}\left [\vec{x}^{(\downarrow)}\right ] \right |
\end{equation}
where:
\begin{equation}
\Phi^{(\alpha)}\left [\vec{x}^{(\alpha)} \right ] =
\begin{bmatrix} 
\phi^{(\alpha)}_0(x_0) & \phi^{(\alpha)}_1(x_0) & \cdots & \phi^{(\alpha)}_{N^{(\alpha)}-1}(x_0) \\
\phi^{(\alpha)}_0(x_1) & \phi^{(\alpha)}_1(x_1) & \cdots & \phi^{(\alpha)}_{N^{(\alpha)}-1}(x_1) \\
\vdots & \vdots & \ddots & \vdots \\
\phi^{(\alpha)}_0(x_{N^{(\alpha)}-1}) & \phi^{(\alpha)}_1(x_{N^{(\alpha)}-1}) & \cdots & \phi^{(\alpha)}_{N^{(\alpha)}-1}(x_{N^{(\alpha)}-1})
\end{bmatrix}
\label{eq:orbmat}.
\end{equation}
If $\Phi^{(\alpha)}$ is chosen to be spin-independent, the ansatz reduces to the restricted Hartree–Fock (RHF) state, which was employed in our previous simulations.

For case of $N^{(\uparrow)}=N^{(\downarrow)}$, we may alternatively employ a BCS-like wave function of the form $\ket{\Psi_\text{bcs}}=(F_{kk'}c_{k\uparrow}^\dagger c_{k'\downarrow}^\dagger)^{N/2}$, where $F$ is the pairing matrix and $c_k^\dagger$ creates an electron in the plane-wave basis.
In practice, however, the dimension of the plane-wave basis is much larger than the number of electrons, rendering a full parameterization of the pairing matrix $F_{ij}$ inefficient.
To reduce the number of variational parameters, we first construct a reduced set of $n_\text{orb}>N/2$ single-particle orbitals from the plane-wave basis, $c^\dagger_{i\alpha} = \sum_{\vec k} A^{\alpha}_{i,\vec k}c^\dagger_{\vec{k}}$,
where $i$ labels the orbital index and $A^\alpha$ are variational orbital-parameter matrices.
The BCS-like wave function can then be written as $\ket{\Psi_\text{bcs}}=(\sum_i S_ic_{i\uparrow}^\dagger c_{i'\downarrow}^\dagger)^{N/2}$,
where $S_i$ are diagonal pairing parameters. This form is sufficient because any non-diagonal structure of the pairing matrix can be absorbed into the orbital rotations encoded in $A^\alpha$, leaving only the diagonal pairing amplitudes $S_i$ as independent variational parameters.
In our simulations, we typically choose the orbital dimension to be $n_\text{orb}=N^{\uparrow}+N^{\downarrow}$.

To be more explicit, in real-space coordinates $\vec x$ the BCS-like wave function can be evaluated as
\begin{equation}
\label{eqn:BCS}
    \Psi(\vec{x}) = \det\left [ \Phi^{(\uparrow)}\left [\vec{x}^{(\uparrow)}\right ] S ({\Phi^{(\downarrow)}}\left [\vec{x}^{(\downarrow)}\right ])^T \right ],
\end{equation}
where $\Phi^\alpha \in \mathbb C^{N^\alpha, n_\text{orb}}$ are the spin-resolved orbital matrices, defined as in Eq.~\eqref{eq:orbmat} but with an enlarged orbital dimension $n_\text{orb}>N^\alpha$.

\begin{figure}
    \centering
    \includegraphics[width=0.5\linewidth]{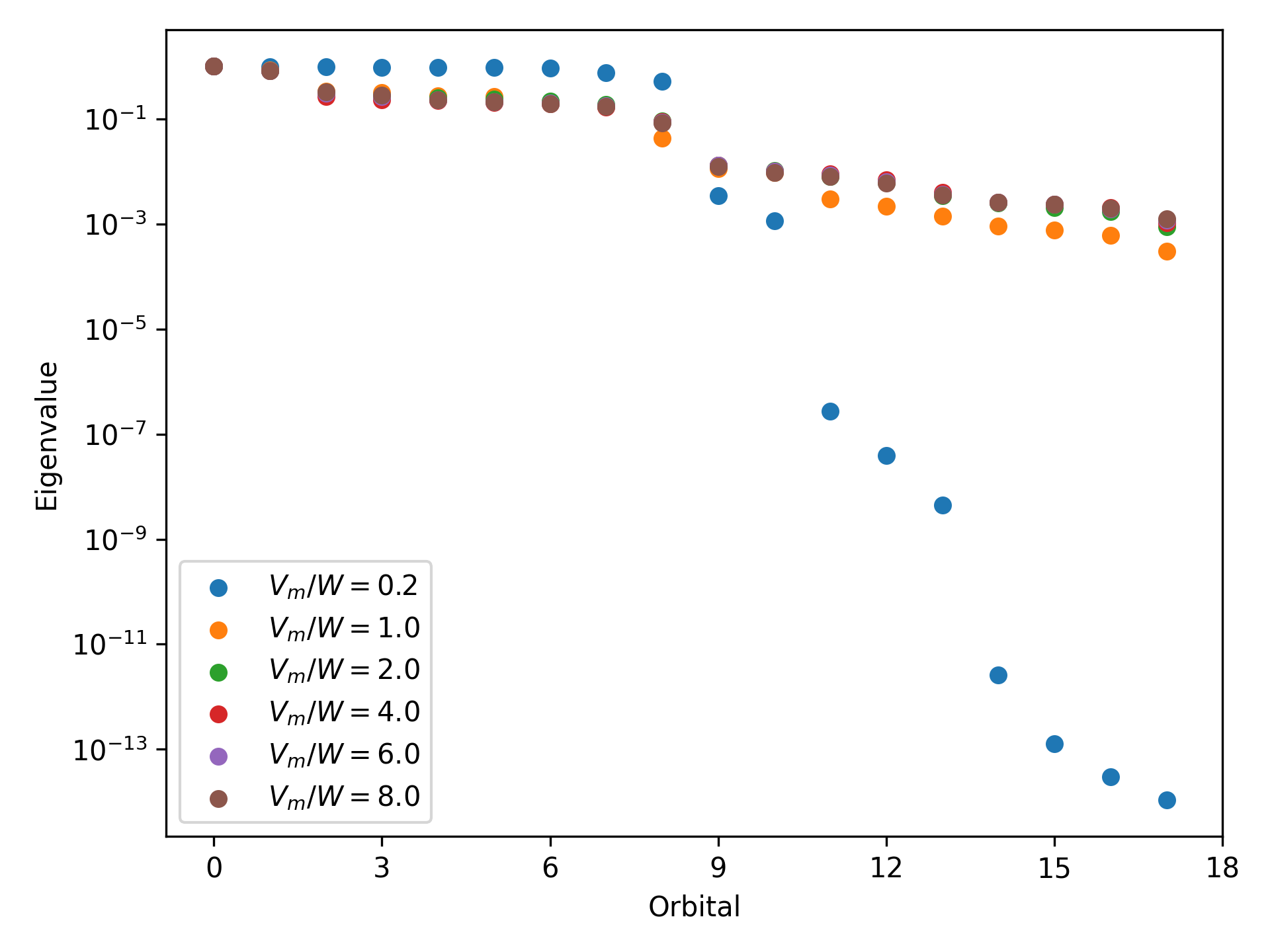}
    \caption{Eigenvalues of orbitals in the projected-BCS wavefunction}
    \label{fig:BCS_eigs}
\end{figure}

Both the standard form in Eq.~\ref{eqn:spin_split} and the BCS form identify the PWC as the lowest-energy state. 
However, the BCS wave function consistently yields a lower variational energy. 
A detailed comparison of the energies is presented in Table~\ref{tab:polar_moire} where we also include the energy of the fully polarized system to illustrate the energy lowering from pairing electrons in singlets.
To better illustrate the structure encoded by the BCS wave function, we compute the occupation amplitudes, defined as the singular values of the effective pairing matrix in the present parameterization,
$F\equiv{A^{(\downarrow)}}^TSA^{(\uparrow)}$.
These occupation amplitudes generally differ from the diagonal matrix $S$ because no orthogonality constraint is imposed on the orbital operators $c_{i,\alpha}^\dagger$.
In Fig.~\ref{fig:BCS_eigs}, we plot the occupation amplitudes for different values of $V_m/W$. 
For $V_m/W=0.1$, where the ground state is expected to be well captured by an RHF–like wave function, the occupation amplitudes exhibit a sharp drop at the $N/2$-th largest value, indicating that the state effectively reduces to an RHF state.
As $V_m/W$ is increased and the paired Wigner crystal phase emerges, the occupation amplitudes display qualitatively different behavior compared to the $V_m/W=0.1$. In particular, the tail of the occupation spectrum persists and becomes increasingly pronounced with increasing $V_m/W$. 
This behavior provides evidence that the pairing mechanism encoded in the BCS wave function plays an essential role in stabilizing the paired Wigner crystal phase, consistent with the observed lowering of the variational energy.

\section{Comparing Energies of Candidate Ground States}\label{sec:energetics_moire}
It is always possible that there is a state lower in energy that is qualitatively different than the PWC. However, in this section we provide more evidence supporting the fact that it is the ground state by comparing with various candidate states and providing energies from DMC.

\begin{table}[!htb]
\centering
\begin{tabular}{|l|l|}
\hline
           & Mean       \\ \hline
NN-VMC     & -2.2865(2) \\ \hline
NN-BCS-VMC & -2.2999(2)   \\ \hline
LDA-DMC    & -2.2991(6) \\ \hline
NNOrb-DMC  & -2.3007(2) \\ \hline
\end{tabular}
\caption{Energies for $V_m/W=2$ comparing Variational Monte Carlo with Neural Network backflow to DMC with orbitals of LDA, Slater-Jastrow and those found by the neural network.}\label{tab:orb_compare}
\end{table}
In Table~\ref{tab:orb_compare}, we compare the energies obtained with the neural network with and without the BCS determinant to those obtained with DMC. The LDA orbitals do not find the PWC but find comparable energies to the BCS \ansatz\ when the LDA orbitals are used in a DMC calculation. To ensure that the PWC we find is not explicitly the result of a bias in our ansatz, we test DMC with orbitals obtained from the \ansatz. Both cases are lower than LDA-DMC and lower than the \ansatz\ and find the PWC and so we are confident this is not a result of a bias in the optimization of our ansatz. It is likely that if we run DMC on the full network and orbitals, the energy will lower more but that is left for future work.

\begin{table}[!htb]
\centering
\begin{tabular}{|l|l|l|l|l|}
\hline
          & $V_m/W=2.0$      & $V_m/W=4.0$     & $V_m/W=6.0$      & $V_m/W=8.0$      \\ \hline
NN-P      & -1.77211(6) & -2.2833(2) & -2.88069(7) & -3.53534(9) \\ \hline
NN-UP     & -1.7910(1)  & -2.2865(2) & -2.8882(2)  & -3.5482(2)  \\ \hline
NN-BCS-UP & -1.7953(4)  & -2.2999(2) & -2.8955(5)  & -3.5515(2)  \\ \hline
\end{tabular}
\caption{Ground state energies comparing Variational Monte Carlo with Neural Network backflow for the polarized system, the unpolarized system with a standard determinant and the unpolarized with a BCS type determinant.}\label{tab:polar_moire}
\end{table}
\COMMENTED{
\begin{table}[!htb]
\begin{tabular}{|l|l|l|l|l|}
\hline
          & $V_m=2.0$      & $V_m=4.0$     & $V_m=6.0$      & $V_m=8.0$      \\ \hline
NN-P      & -1.77211(7) & -2.2833(2) & -2.88069(7) & -3.53534(9) \\ \hline
NN-UP     & -1.7905(3)  & -2.2866(2) & -2.8883(2)  & -3.5482(1)  \\ \hline
NN-BCS-UP & -1.7953(4)  & -2.2997(3) & -2.8974(6)  & -3.5516(2)  \\ \hline
NN-Pre-UP & None  & -2.2989 & None  & None  \\ \hline
NN-Pre-UP-BCS &  None  & -2.2973 & None  & None  \\ \hline
\end{tabular}
\cs{Put these in a plot. NN-Pre-UP includes the prefactor in the orbitals similar to the magnetic field ansatz used for the rotating MWCs. In contrast to that case, this includes backflow and prefactors from a single network - should remove this result here}
\caption{Ground state energies comparing Variational Monte Carlo with Neural Network backflow for the polarized system, the unpolarized system with a standard determinant and the unpolarized with a BCS type determinant.}\label{tab:polar_moire}
\end{table}
}

\section{Paired Wigner Crystal in Different Lattices, Geometries and Densities}
\begin{figure}[!htbp]
  \centering

  \begin{subfigure}[t]{0.24\textwidth}
    \centering
    \includegraphics[width=\linewidth,height=0.85\textheight,keepaspectratio]{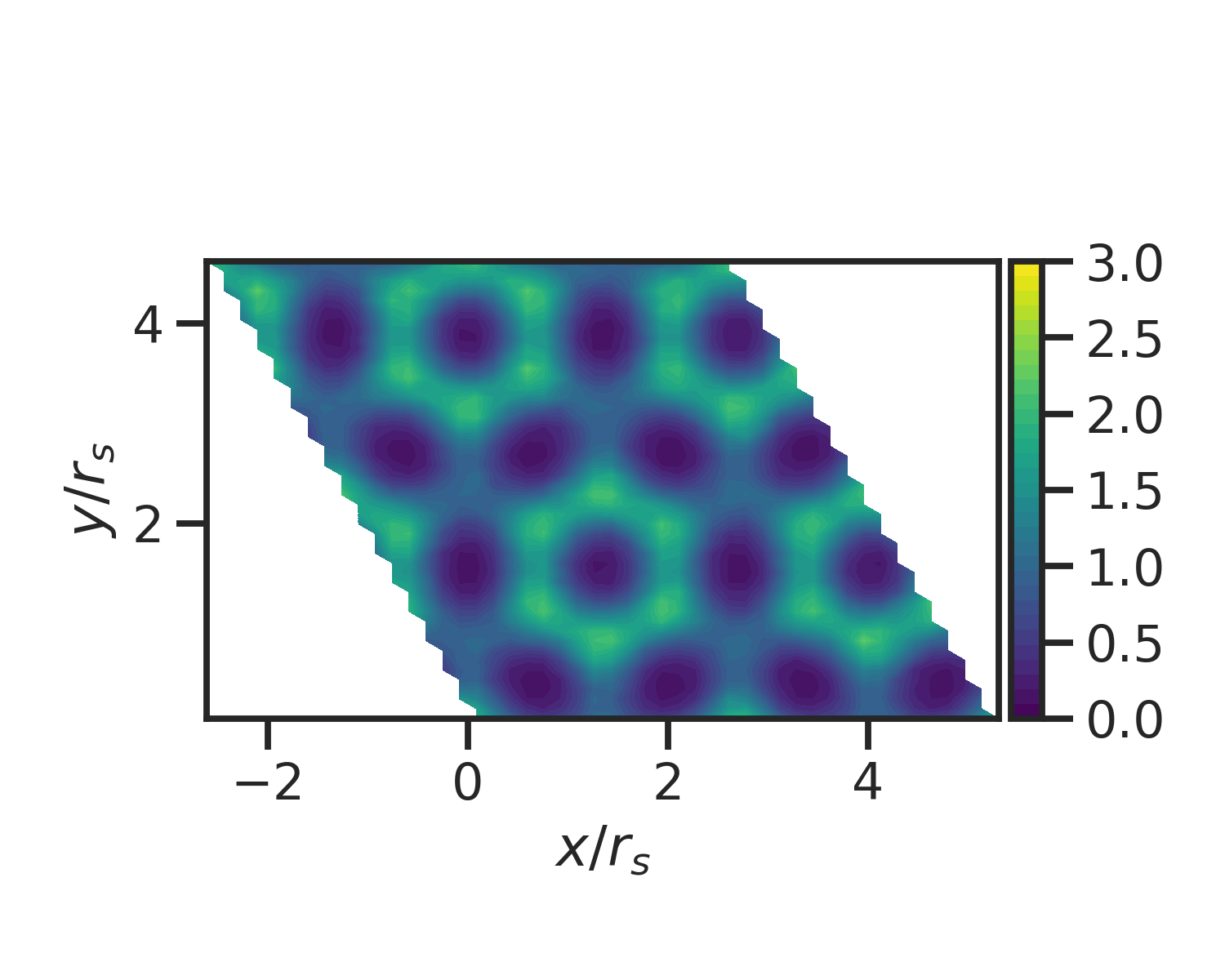}
    \caption{}
    \label{fig:size_comp:a}
  \end{subfigure}\hfill
  \begin{subfigure}[t]{0.24\textwidth}
    \centering
    \includegraphics[width=\linewidth,height=0.85\textheight,keepaspectratio]{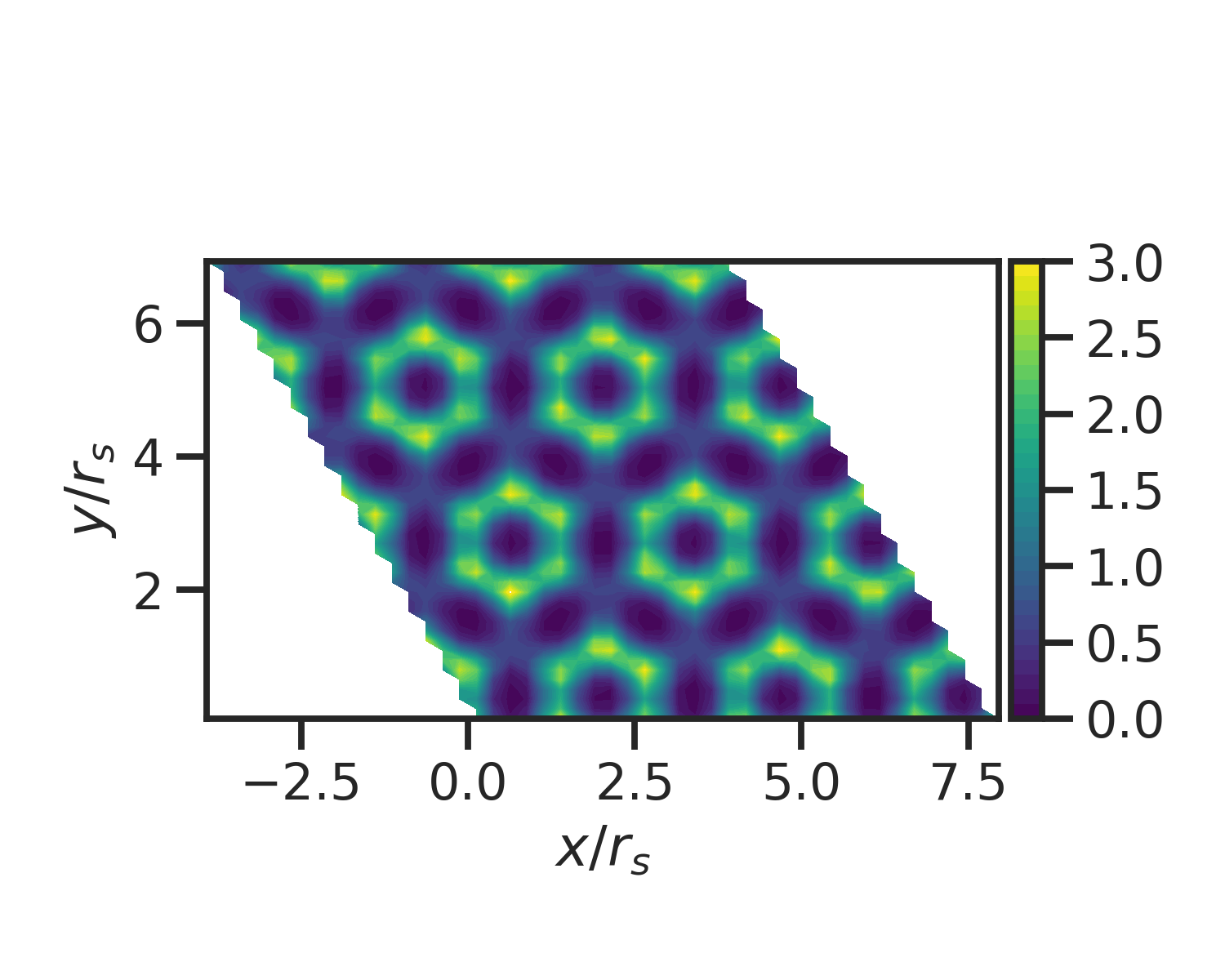}
    \caption{}
    \label{fig:size_comp:b}
  \end{subfigure}\hfill
  \begin{subfigure}[t]{0.24\textwidth}
    \centering
    \includegraphics[width=\linewidth,height=0.85\textheight,keepaspectratio]{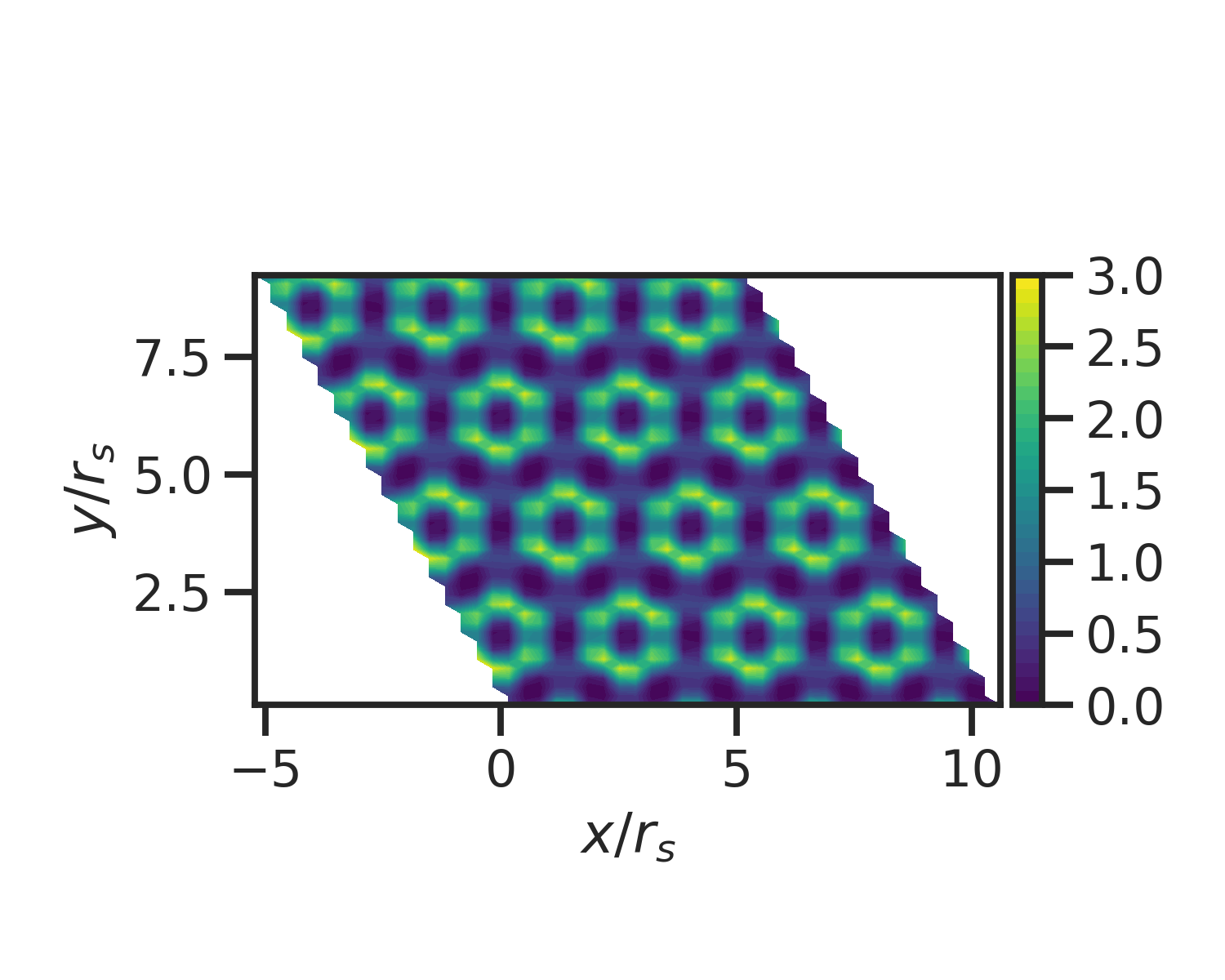}
    \caption{}
    \label{fig:size_comp:c}
  \end{subfigure}
  \begin{subfigure}[t]{0.2\textwidth}
    \centering
    \includegraphics[width=\linewidth,height=0.85\textheight,keepaspectratio]{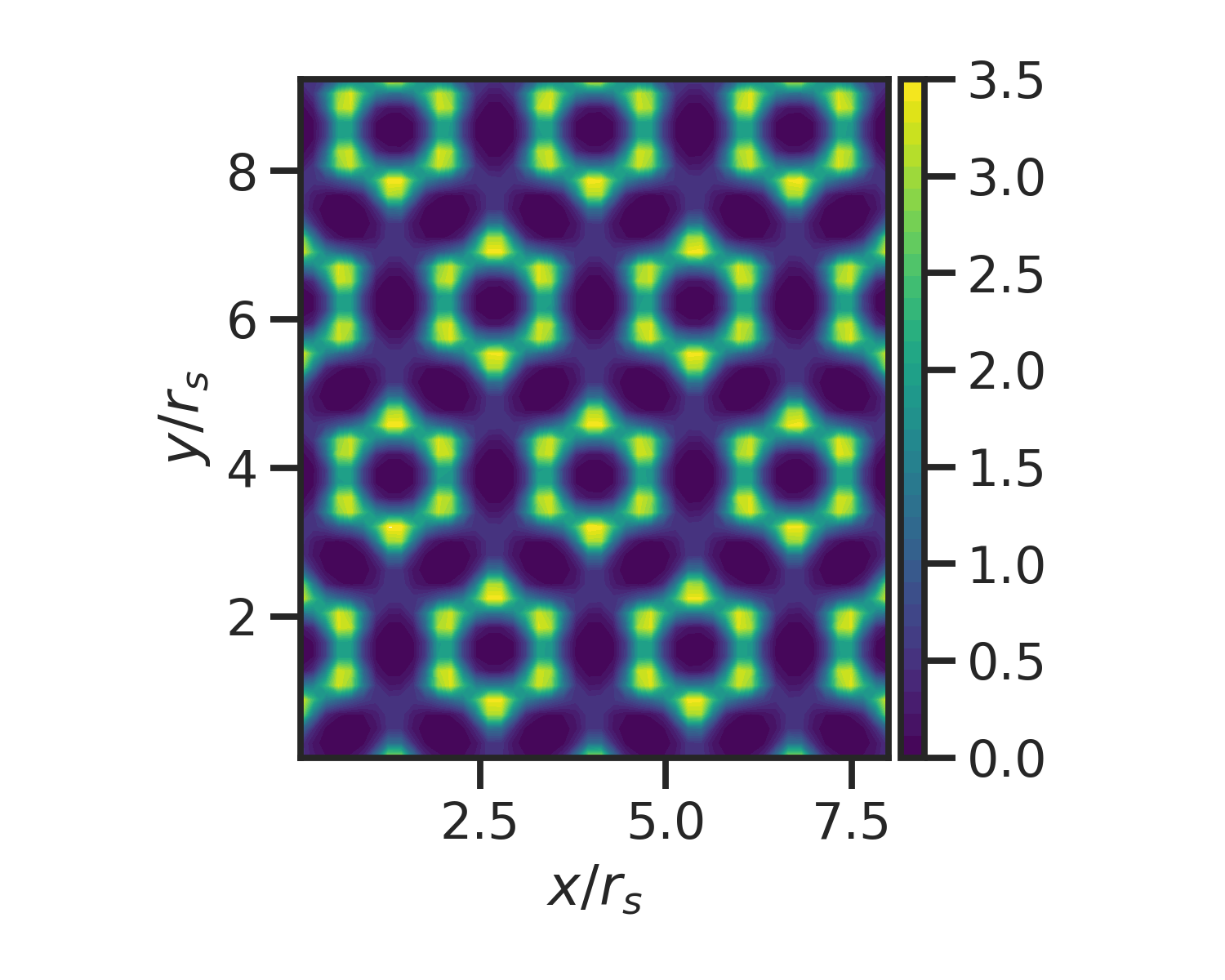}
    \caption{}
    \label{fig:size_comp:d}
  \end{subfigure}

  \caption{Ground state charge densities are shown for $r_s=10$ with lattice sizes and potential depths of (a) $4\times4$, $V_m/W=2.0$, (b) $6\times6$, $V_m/W=4.0$, (c) $8\times8$, $V_m/W=6.0$ and (d) $6\times 8$, $V_m/W=8.0$. All simulations are done at the same filling, $\nu=1/4$, as in the main text. (a-c) are in a triangular cell while (d) is in a rectangular cell.}
  \label{fig:size_comp}
\end{figure}

To check that the PWC is not an artifact of cell size or shape, we show various system sizes and compare the triangular cell to the rectangular cell. All cases utilize the same filling of $\nu=1/4$. This corresponds to $8$ electrons on the $4\times4$ lattice, $18$ electrons on the $6\times6$ lattice, 32 electrons on the $8 \times 8$ lattice and $24$ electrons on the $6\times 8$ lattice. From the charge density plots in Fig.~\ref{fig:size_comp} it is clear that the PWC is robust from a $4\times4$ lattice to an $8\times8$ lattice and in triangular as well as rectangular cells.

\begin{figure}[!htbp]
  \centering

  \begin{subfigure}[t]{0.48\textwidth}
    \centering
    \includegraphics[width=\linewidth,height=0.85\textheight,keepaspectratio]{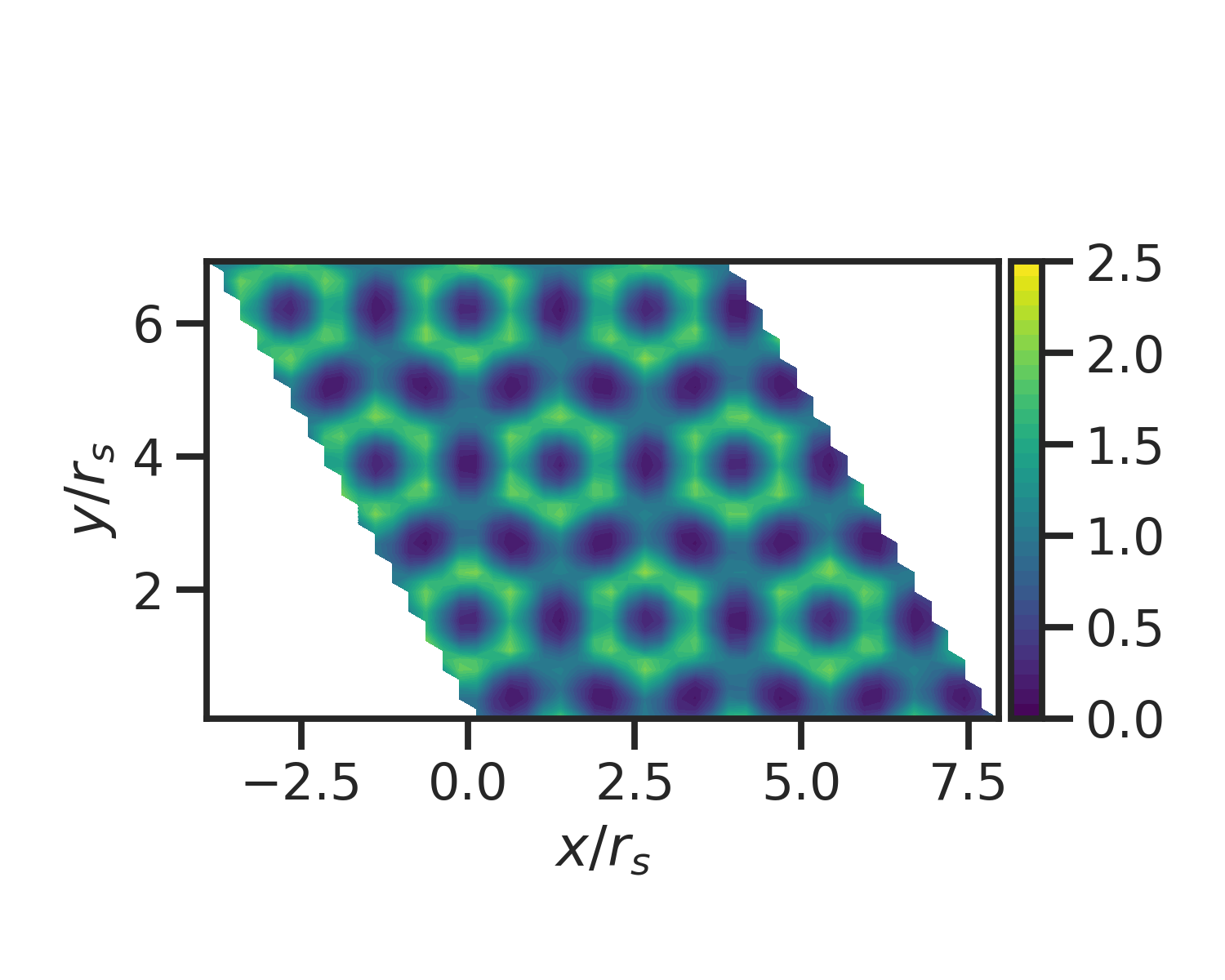}
    \caption{}
    \label{fig:rs_comp:a}
  \end{subfigure}\hfill
  \begin{subfigure}[t]{0.48\textwidth}
    \centering
    \includegraphics[width=\linewidth,height=0.85\textheight,keepaspectratio]{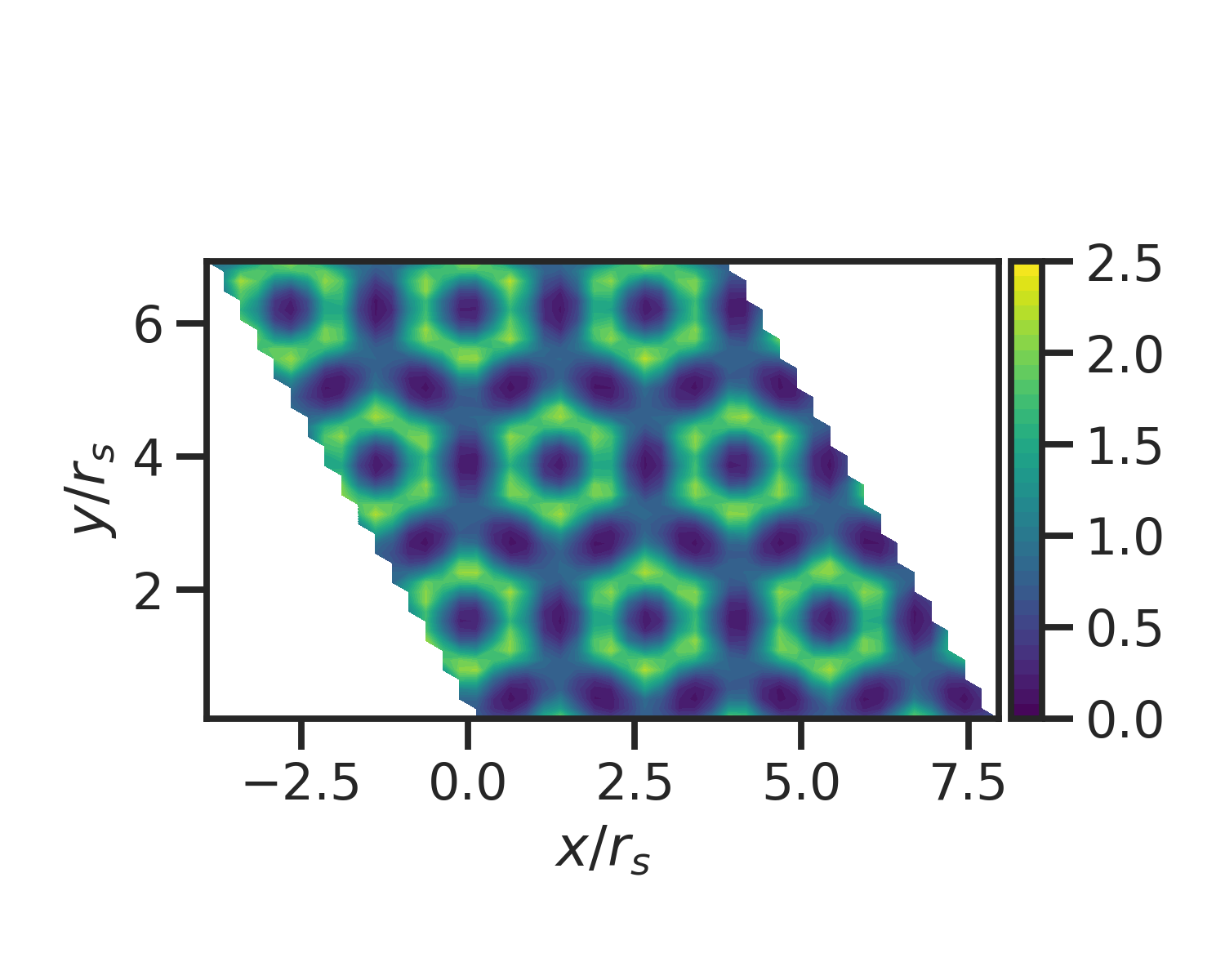}
    \caption{}
    \label{fig:rs_comp:b}
  \end{subfigure}\hfill

  \caption{Ground state charge densities are shown for $V_m=2.0$ at $r_s=5$ in (a) and $r_s=15$ in (b). Both simulations are done at the same filling, $\nu=1/4$, as in the main text in a triangular cell on a $6\times 6$ lattice.}
  \label{fig:rs_comp}
\end{figure}

To show that this phase is robust over a region accessible by various experimental devices, we also include here the result at $r_s=5$ and $r_s=15$ whereas the results in the main text are strictly at $r_s=10$. In Fig.~\ref{fig:rs_comp} we show charge densities at $V_M/W=2.0$ for $r_s=5$ and $r_s=15$. Both systems are on a $6\times6$ lattice with 18 electrons and exhibit the same hexagonal PWC pattern. In sum, this gives us confidence that the phase is robust to cell shape, is not a finite size effect and is stable over a density of at least $r_s=5$ to $r_s=15$.

\section{Descriptions of the Density Functional Theory and Diffusion Monte Carlo Calculations}\label{sec:supp_DMC}

In addition to the NQS calculations, we also looked for solutions of the moir\'e continuum Hamiltonian Eq.~(\ref{eq:mch}) using traditional electronic structure methods including density functional theory (DFT) and diffusion Monte Carlo (DMC). The numerical details are consistent with those in Refs.~\cite{Yang_MetalInsulatorTransitionSemiconductor_2024,Yang_FerromagneticSemimetalChargeDensity_2024}, but with the system modified to match that explored by the NQS calculations: $18$ electrons in $36$ moir\'e unit cells with periodic boundary conditions.

\begin{figure*}[ht]
\centering
\begin{minipage}{0.32\textwidth}
\includegraphics[width=\linewidth]{{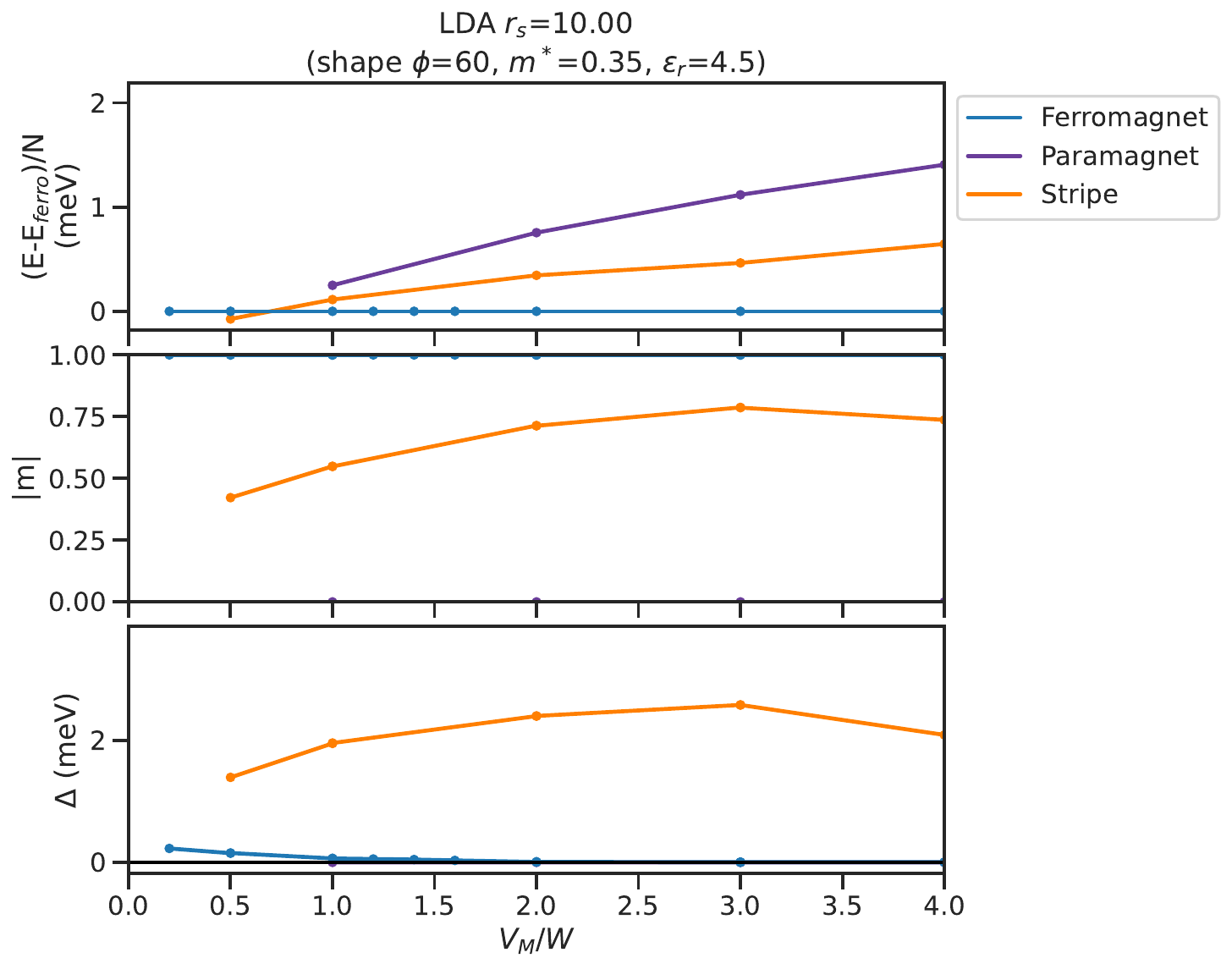}}
(a) LDA
\end{minipage}
\begin{minipage}{0.32\textwidth}
\includegraphics[width=\linewidth]{{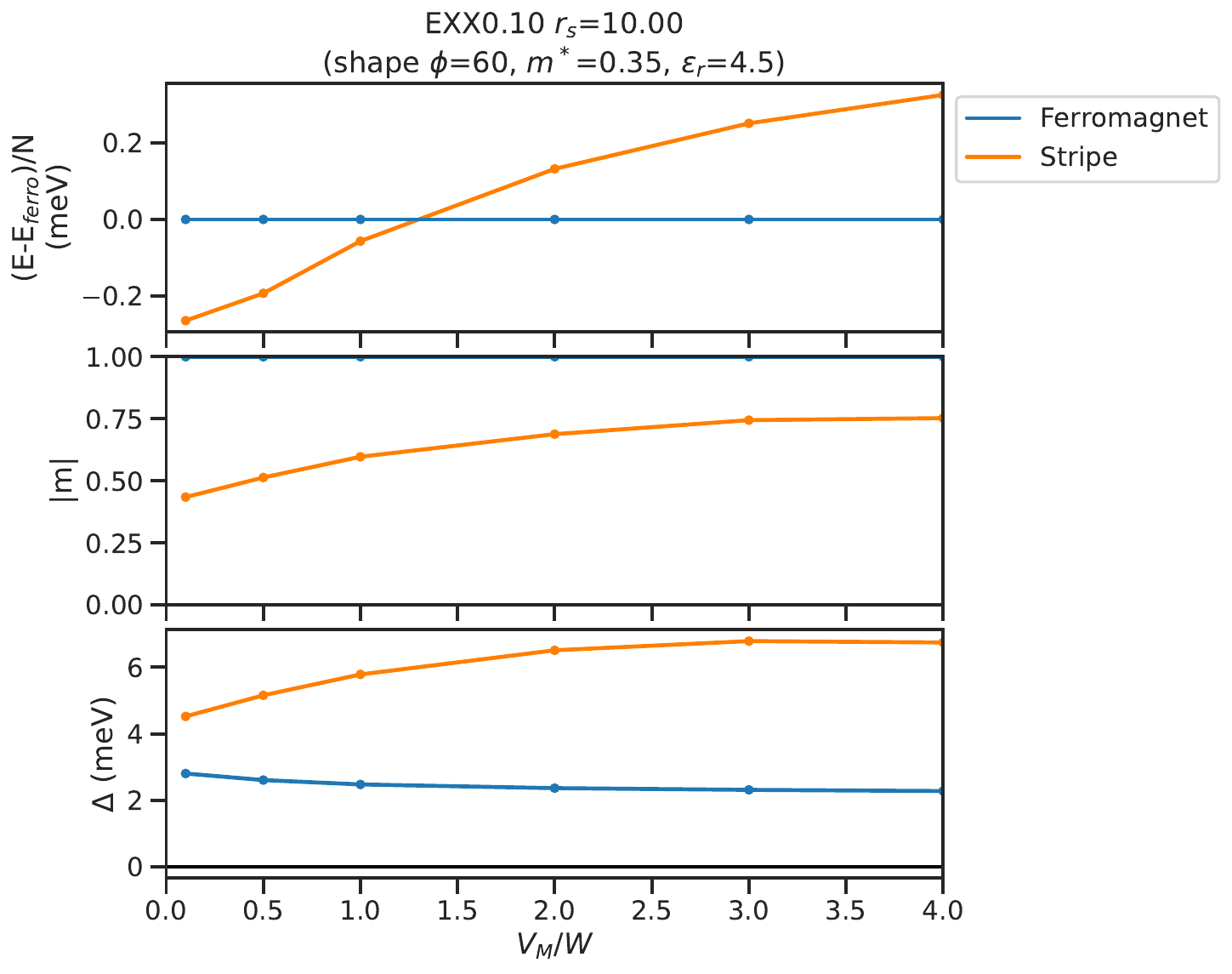}}
(b) LDA + $10\%$ exact exchange
\end{minipage}
\begin{minipage}{0.32\textwidth}
\includegraphics[width=\linewidth]{{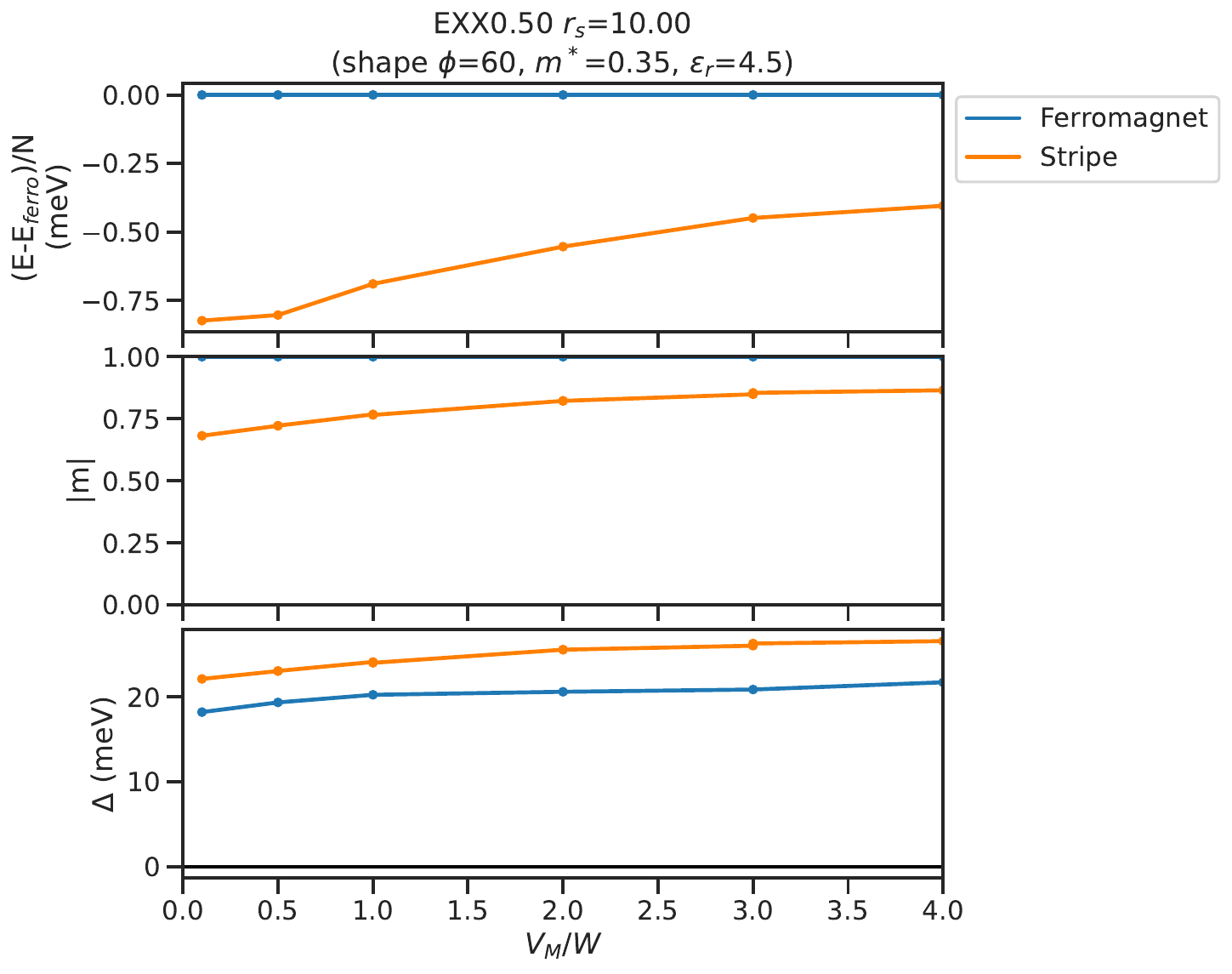}}
(c) LDA + $50\%$ exact exchange
\end{minipage}
\caption{DFT energy, magnetization, and gap as functions of the depth of the moir\'e potential at $r_s=10$.}
\label{fig:dft-rs10}
\end{figure*}

\begin{figure}[h]
\centering
\includegraphics[width=0.9\linewidth]{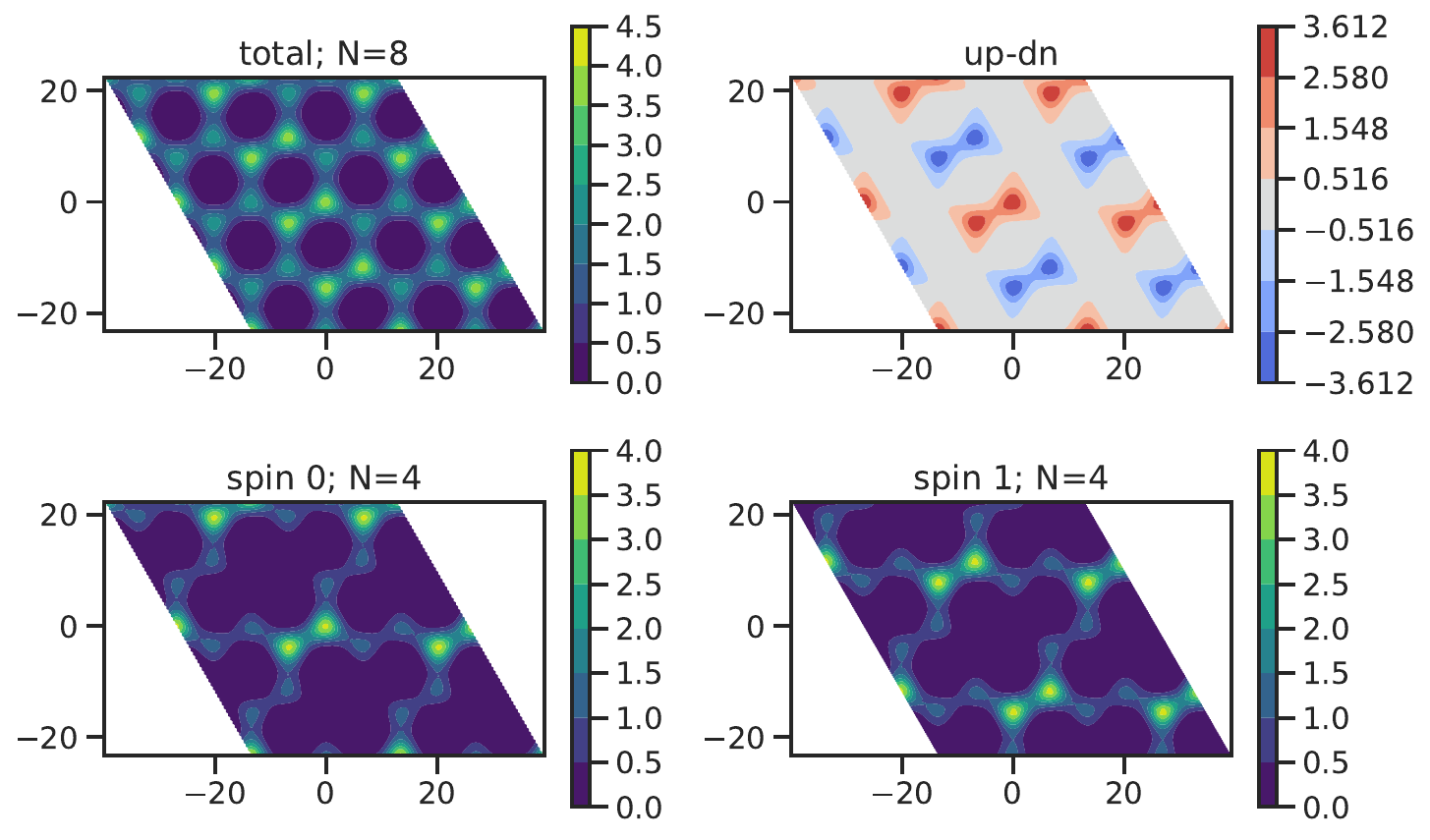}
\caption{Charge and spin densities of the antiferromagnetic DFT state that best approximates the paired Wigner crystal.}
\label{fig:dft_state}
\end{figure}

We were unable to locate the ring state using DFT.
As shown in Fig.~\ref{fig:dft-rs10}, LDA predicts a gapless ferromagnetic ground state starting in moderately deep moir\'e potential $V_M/W\ge1$ which persists in deeper potentials.
The introduction of exact exchange into LDA stabilizes an anti-ferromagnetic state.
However, instead of singlets that span hexagonal rings, this state can be viewed as two staggered triangular lattices, one formed by up spins, while the other by down spins.
As shown in Fig.~\ref{fig:dft_state}, each electron is delocalized across two moir\'e minima, forming a ``bowtie'' pattern. Opposite-spin electrons occupy opposing edges of what would have been honeycomb rings from the NQS calculation.
As such, this state can be viewed as the ``closest'' mean-field solution that approximates the paired Wigner crystal. %

To obtain the energies presented in Table~\ref{tab:orb_compare}, we performed DMC calculations using Slater-Jastrow trial wavefunctions using LDA and neural-network orbitals. For each set of orbitals, we optimized short-range two-body Jastrows and used the resultant trial wavefunctions in subsequent DMC calculations with timesteps of $0.25$ ha$^{-1}$, which resulted in an acceptance ratio of $99.8\%$. We used $1024$ walkers, which is more than sufficient for this small system of $N=36$ electrons.

\twocolumngrid
\bibliographystyle{apsrev4-2}
\bibliography{refs}

@article{Yang_FerromagneticSemimetalChargeDensity_2024,
  title = {Ferromagnetic {{Semimetal}} and {{Charge-Density Wave Phases}} of {{Interacting Electrons}} in a {{Honeycomb Moir{\'e} Potential}}},
  author = {Yang, Yubo and Morales, Miguel A. and Zhang, Shiwei},
  year = {2024},
  month = dec,
  journal = {Phys. Rev. Lett.},
  volume = {133},
  number = {26},
  pages = {266501},
  issn = {0031-9007, 1079-7114},
  doi = {10.1103/PhysRevLett.133.266501},
  urldate = {2025-03-19},
  langid = {english}
}

@article{Yang_MetalInsulatorTransitionSemiconductor_2024,
  title = {Metal-{{Insulator Transition}} in a {{Semiconductor Heterobilayer Model}}},
  author = {Yang, Yubo and Morales, Miguel A. and Zhang, Shiwei},
  year = {2024},
  month = feb,
  journal = {Phys. Rev. Lett.},
  volume = {132},
  number = {7},
  pages = {076503},
  issn = {0031-9007, 1079-7114},
  doi = {10.1103/PhysRevLett.132.076503},
  urldate = {2024-02-24},
  langid = {english}
}

@article{Kennes_MoireHeterostructuresCondensedmatter_2021,
  title = {Moir{\'e} Heterostructures as a Condensed-Matter Quantum Simulator},
  author = {Kennes, Dante M. and Claassen, Martin and Xian, Lede and Georges, Antoine and Millis, Andrew J. and Hone, James and Dean, Cory R. and Basov, D. N. and Pasupathy, Abhay N. and Rubio, Angel},
  year = {2021},
  month = feb,
  journal = {Nat. Phys.},
  volume = {17},
  number = {2},
  pages = {155--163},
  issn = {1745-2473, 1745-2481},
  doi = {10.1038/s41567-020-01154-3},
  urldate = {2023-07-06},
  langid = {english}
}

@article{Mak_SemiconductorMoireMaterials_2022,
  title = {Semiconductor Moir{\'e} Materials},
  author = {Mak, Kin Fai and Shan, Jie},
  year = {2022},
  month = jul,
  journal = {Nat. Nanotechnol.},
  volume = {17},
  number = {7},
  pages = {686--695},
  issn = {1748-3387, 1748-3395},
  doi = {10.1038/s41565-022-01165-6},
  urldate = {2023-07-06},
  langid = {english}
}

@ARTICLE{Ma_RelativisticMott_2024,
       author = {{Ma}, Liguo and {Chaturvedi}, Raghav and {Nguyen}, Phuong X. and {Watanabe}, Kenji and {Taniguchi}, Takashi and {Mak}, Kin Fai and {Shan}, Jie},
        title = "{Relativistic Mott transition in strongly correlated artificial graphene}",
      journal = {arXiv e-prints},
         year = 2024,
        month = dec,
          eid = {arXiv:2412.07150},
        pages = {arXiv:2412.07150},
          doi = {10.48550/arXiv.2412.07150},
archivePrefix = {arXiv},
       eprint = {2412.07150},
 primaryClass = {cond-mat.mes-hall},
       adsurl = {https://ui.adsabs.harvard.edu/abs/2024arXiv241207150M}
}

@article{Adak_TunableMoireMaterials_2024a,
  title = {Tunable Moir{\'e} Materials for Probing {{Berry}} Physics and Topology},
  author = {Adak, Pratap Chandra and Sinha, Subhajit and Agarwal, Amit and Deshmukh, Mandar M.},
  year = {2024},
  month = apr,
  journal = {Nat Rev Mater},
  volume = {9},
  number = {7},
  pages = {481--498},
  issn = {2058-8437},
  doi = {10.1038/s41578-024-00671-4},
  urldate = {2025-06-23},
  langid = {english}
}

@article{Andrei_MarvelsMoireMaterials_2021,
  title = {The Marvels of Moir{\'e} Materials},
  author = {Andrei, Eva Y. and Efetov, Dmitri K. and {Jarillo-Herrero}, Pablo and MacDonald, Allan H. and Mak, Kin Fai and Senthil, T. and Tutuc, Emanuel and Yazdani, Ali and Young, Andrea F.},
  year = {2021},
  month = mar,
  journal = {Nat Rev Mater},
  volume = {6},
  number = {3},
  pages = {201--206},
  issn = {2058-8437},
  doi = {10.1038/s41578-021-00284-1},
  urldate = {2023-07-31},
  langid = {english}
}

@article{Checkelsky_FlatBandsStrange_2024a,
  title = {Flat Bands, Strange Metals and the {{Kondo}} Effect},
  author = {Checkelsky, Joseph G. and Bernevig, B. Andrei and Coleman, Piers and Si, Qimiao and Paschen, Silke},
  year = {2024},
  month = feb,
  journal = {Nat Rev Mater},
  volume = {9},
  number = {7},
  pages = {509--526},
  issn = {2058-8437},
  doi = {10.1038/s41578-023-00644-z},
  urldate = {2025-06-23},
  langid = {english}
}

@article{Nuckolls_MicroscopicPerspectiveMoire_2024,
  title = {A Microscopic Perspective on Moir{\'e} Materials},
  author = {Nuckolls, Kevin P. and Yazdani, Ali},
  year = {2024},
  month = may,
  journal = {Nat Rev Mater},
  volume = {9},
  number = {7},
  pages = {460--480},
  issn = {2058-8437},
  doi = {10.1038/s41578-024-00682-1},
  urldate = {2025-06-05},
  langid = {english}
}

@ARTICLE{Zhou_Electroinc_2025,
       author = {{Zhou}, You and {Esterlis}, Ilya and {Smole{\'n}ski}, Tomasz},
        title = "{Electronic crystals in layered materials}",
      journal = {arXiv e-prints},
         year = 2025,
        month = sep,
          eid = {arXiv:2509.21222},
        pages = {arXiv:2509.21222},
          doi = {10.48550/arXiv.2509.21222},
archivePrefix = {arXiv},
       eprint = {2509.21222},
 primaryClass = {cond-mat.str-el},
       adsurl = {https://ui.adsabs.harvard.edu/abs/2025arXiv250921222Z}
}

@article{wang_correlated_2020,
  title = {Correlated Electronic Phases in Twisted Bilayer Transition Metal Dichalcogenides},
  author = {Wang, Lei and Shih, En-Min and Ghiotto, Augusto and Xian, Lede and Rhodes, Daniel A. and Tan, Cheng and Claassen, Martin and Kennes, Dante M. and Bai, Yusong and Kim, Bumho and Watanabe, Kenji and Taniguchi, Takashi and Zhu, Xiaoyang and Hone, James and Rubio, Angel and Pasupathy, Abhay N. and Dean, Cory R.},
  year = {2020},
  month = aug,
  journal = {Nature Materials},
  volume = {19},
  number = {8},
  pages = {861--866},
  issn = {1476-1122, 1476-4660},
  doi = {10.1038/s41563-020-0708-6},
  urldate = {2023-07-06},
  langid = {english}
}

@article{regan_mott_2020,
  title = {Mott and Generalized {{Wigner}} Crystal States in {{WSe2}}/{{WS2}} Moir{\'e} Superlattices},
  author = {Regan, Emma C. and Wang, Danqing and Jin, Chenhao and Bakti Utama, M. Iqbal and Gao, Beini and Wei, Xin and Zhao, Sihan and Zhao, Wenyu and Zhang, Zuocheng and Yumigeta, Kentaro and Blei, Mark and Carlstr{\"o}m, Johan D. and Watanabe, Kenji and Taniguchi, Takashi and Tongay, Sefaattin and Crommie, Michael and Zettl, Alex and Wang, Feng},
  year = {2020},
  month = mar,
  journal = {Nature},
  volume = {579},
  number = {7799},
  pages = {359--363},
  issn = {0028-0836, 1476-4687},
  doi = {10.1038/s41586-020-2092-4},
  urldate = {2023-07-06},
  langid = {english}
}

@article{Li_Continuous_2021,
arxivId = {2103.09779},
author = {Li, Tingxin and Jiang, Shengwei and Li, Lizhong and Zhang, Yang and Kang, Kaifei and Zhu, Jiacheng and Watanabe, Kenji and Taniguchi, Takashi and Chowdhury, Debanjan and Fu, Liang and Shan, Jie and Mak, Kin Fai},
doi = {10.1038/s41586-021-03853-0},
eprint = {2103.09779},
issn = {0028-0836},
journal = {Nature},
month = {sep},
number = {7876},
pages = {350--354},
publisher = {Springer US},
title = {{Continuous Mott transition in semiconductor moiré superlattices}},
url = {http://arxiv.org/abs/2103.09779 https://www.nature.com/articles/s41586-021-03853-0},
volume = {597},
year = {2021}
}

@article{shabani_deep_2021,
  title = {Deep Moir{\'e} Potentials in Twisted Transition Metal Dichalcogenide Bilayers},
  author = {Shabani, Sara and Halbertal, Dorri and Wu, Wenjing and Chen, Mingxing and Liu, Song and Hone, James and Yao, Wang and Basov, D. N. and Zhu, Xiaoyang and Pasupathy, Abhay N.},
  year = {2021},
  month = jun,
  journal = {Nature Physics},
  volume = {17},
  number = {6},
  pages = {720--725},
  issn = {1745-2473, 1745-2481},
  doi = {10.1038/s41567-021-01174-7},
  urldate = {2023-07-06},
  langid = {english}
}

@article{nieken_direct_2022,
  title = {Direct {{STM}} Measurements of {{R-type}} and {{H-type}} Twisted {{MoSe}} {\textsubscript{2}} /{{WSe}} {\textsubscript{2}}},
  author = {Nieken, Rachel and Roche, Anna and Mahdikhanysarvejahany, Fateme and Taniguchi, Takashi and Watanabe, Kenji and Koehler, Michael R. and Mandrus, David G. and Schaibley, John and LeRoy, Brian J.},
  year = {2022},
  month = mar,
  journal = {APL Materials},
  volume = {10},
  number = {3},
  pages = {031107},
  issn = {2166-532X},
  doi = {10.1063/5.0084358},
  urldate = {2023-07-06},
  langid = {english}
}

@article{cai_signatures_2023,
  title = {Signatures of Fractional Quantum Anomalous {{Hall}} States in Twisted {{MoTe2}}},
  author = {Cai, Jiaqi and Anderson, Eric and Wang, Chong and Zhang, Xiaowei and Liu, Xiaoyu and Holtzmann, William and Zhang, Yinong and Fan, Fengren and Taniguchi, Takashi and Watanabe, Kenji and Ran, Ying and Cao, Ting and Fu, Liang and Xiao, Di and Yao, Wang and Xu, Xiaodong},
  year = {2023},
  month = oct,
  journal = {Nature},
  volume = {622},
  number = {7981},
  pages = {63--68},
  issn = {0028-0836, 1476-4687},
  doi = {10.1038/s41586-023-06289-w},
  urldate = {2023-10-07},
  langid = {english}
}

@article{wang_fractional_2024,
  title = {Fractional {{Chern Insulator}} in {{Twisted Bilayer MoTe}} 2},
  author = {Wang, Chong and Zhang, Xiao-Wei and Liu, Xiaoyu and He, Yuchi and Xu, Xiaodong and Ran, Ying and Cao, Ting and Xiao, Di},
  year = {2024},
  month = jan,
  journal = {Physical Review Letters},
  volume = {132},
  number = {3},
  pages = {036501},
  issn = {0031-9007, 1079-7114},
  doi = {10.1103/PhysRevLett.132.036501},
  urldate = {2024-07-17},
  langid = {english},
}

@article{Guo_Superconductivity50degTwisted_2025,
  title = {Superconductivity in 5.0{$^\circ$} Twisted Bilayer {{WSe2}}},
  author = {Guo, Yinjie and Pack, Jordan and Swann, Joshua and Holtzman, Luke and Cothrine, Matthew and Watanabe, Kenji and Taniguchi, Takashi and Mandrus, David G. and Barmak, Katayun and Hone, James and Millis, Andrew J. and Pasupathy, Abhay and Dean, Cory R.},
  year = {2025},
  month = jan,
  journal = {Nature},
  volume = {637},
  number = {8047},
  pages = {839--845},
  issn = {0028-0836, 1476-4687},
  doi = {10.1038/s41586-024-08381-1},
  urldate = {2025-02-25},
  langid = {english}
}

@article{Xia_SuperconductivityTwistedBilayer_2025,
  title = {Superconductivity in Twisted Bilayer {{WSe2}}},
  author = {Xia, Yiyu and Han, Zhongdong and Watanabe, Kenji and Taniguchi, Takashi and Shan, Jie and Mak, Kin Fai},
  year = {2025},
  month = jan,
  journal = {Nature},
  volume = {637},
  number = {8047},
  pages = {833--838},
  issn = {0028-0836, 1476-4687},
  doi = {10.1038/s41586-024-08116-2},
  urldate = {2025-02-25},
  langid = {english}
}

@article{Jia_AnomalousSuperconductivityTwisted_2025,
  title = {Anomalous Superconductivity in Twisted {{MoTe}}{\textsubscript{2}} Nanojunctions},
  author = {Jia, Yanyu and Song, Tiancheng and Zheng, Zhaoyi Joy and Cheng, Guangming and Uzan, Ayelet J. and Yu, Guo and Tang, Yue and Pollak, Connor J. and Yuan, Fang and Onyszczak, Michael and Watanabe, Kenji and Taniguchi, Takashi and Lei, Shiming and Yao, Nan and Schoop, Leslie M. and Ong, N. P. and Wu, Sanfeng},
  year = {2025},
  month = jan,
  journal = {Sci. Adv.},
  volume = {11},
  number = {5},
  pages = {eadq5712},
  issn = {2375-2548},
  doi = {10.1126/sciadv.adq5712},
  urldate = {2025-06-15},
  langid = {english}
}

@article{Moulopoulos_NewLowDensity_1992,
  title = {New Low Density Phase of Interacting Electrons: {{The}} Paired Electron Crystal},
  shorttitle = {New Low Density Phase of Interacting Electrons},
  author = {Moulopoulos, K. and Ashcroft, N. W.},
  year = {1992},
  month = oct,
  journal = {Phys. Rev. Lett.},
  volume = {69},
  number = {17},
  pages = {2555--2558},
  issn = {0031-9007},
  doi = {10.1103/PhysRevLett.69.2555},
  urldate = {2025-02-24},
  langid = {english}
}

@article{Taut_WignerCrystallizationTwodimensional_2001,
  title = {Wigner Crystallization of a Two-Dimensional Electron Gas in a Magnetic Field: {{Single}} Electrons versus Electron Pairs at the Lattice Sites},
  shorttitle = {Wigner Crystallization of a Two-Dimensional Electron Gas in a Magnetic Field},
  author = {Taut, M.},
  year = {2001},
  month = oct,
  journal = {Phys. Rev. B},
  volume = {64},
  number = {16},
  pages = {165315},
  issn = {0163-1829, 1095-3795},
  doi = {10.1103/PhysRevB.64.165315},
  urldate = {2023-07-06},
  langid = {english}
}

@article{Drummond_PhaseDiagramLowDensity_2009,
  title = {Phase {{Diagram}} of the {{Low-Density Two-Dimensional Homogeneous Electron Gas}}},
  author = {Drummond, N. D. and Needs, R. J.},
  year = 2009,
  month = mar,
  journal = {Phys. Rev. Lett.},
  volume = {102},
  number = {12},
  pages = {126402},
  issn = {0031-9007, 1079-7114},
  doi = {10.1103/PhysRevLett.102.126402},
  urldate = {2023-07-06},
  langid = {english}
}

@misc{li2024emergentwignerphasesmoire,
      title={Emergent Wigner phases in moir\'e superlattice from deep learning}, 
      author={Xiang Li and Yubing Qian and Weiluo Ren and Yang Xu and Ji Chen},
      year={2024},
      eprint={2406.11134},
      archivePrefix={arXiv},
      primaryClass={physics.comp-ph},
      url={https://arxiv.org/abs/2406.11134}, 
}

@article{Pescia_MessagepassingNeuralQuantum_2024,
  title = {Message-passing neural quantum states for the homogeneous electron gas},
  author = {Pescia, Gabriel and Nys, Jannes and Kim, Jane and Lovato, Alessandro and Carleo, Giuseppe},
  journal = {Phys. Rev. B},
  volume = {110},
  issue = {3},
  pages = {035108},
  numpages = {11},
  year = {2024},
  month = {Jul},
  publisher = {American Physical Society},
  doi = {10.1103/PhysRevB.110.035108},
  url = {https://link.aps.org/doi/10.1103/PhysRevB.110.035108}
}

@article{Smith_2024,
   title={Unified Variational Approach Description of Ground-State Phases of the Two-Dimensional Electron Gas},
   volume={133},
   ISSN={1079-7114},
   url={http://dx.doi.org/10.1103/PhysRevLett.133.266504},
   DOI={10.1103/physrevlett.133.266504},
   number={26},
   journal={Physical Review Letters},
   publisher={American Physical Society (APS)},
   author={Smith, Conor and Chen, Yixiao and Levy, Ryan and Yang, Yubo and Morales, Miguel A. and Zhang, Shiwei},
   year={2024},
   month=dec }

@article{Angeli2021,
author = {Angeli, Mattia and MacDonald, Allan H.},
doi = {10.1073/pnas.2021826118},
issn = {0027-8424},
journal = {Proceedings of the National Academy of Sciences},
month = {mar},
number = {10},
pages = {1--5},
pmid = {33658375},
title = {{$\Gamma$ valley transition metal dichalcogenide moir{\'{e}} bands}},
url = {https://pnas.org/doi/full/10.1073/pnas.2021826118},
volume = {118},
year = {2021}
}

@article{Resta_ElectronLocalizationInsulating_1999,
  title = {Electron {{Localization}} in the {{Insulating State}}},
  author = {Resta, Raffaele and Sorella, Sandro},
  year = 1999,
  month = jan,
  journal = {Phys. Rev. Lett.},
  volume = {82},
  number = {2},
  pages = {370--373},
  issn = {0031-9007, 1079-7114},
  doi = {10.1103/PhysRevLett.82.370},
  urldate = {2023-07-12},
  langid = {english}
}

@article{Souza_PolarizationLocalizationInsulators_2000,
  title = {Polarization and Localization in Insulators: {{Generating}} Function Approach},
  shorttitle = {Polarization and Localization in Insulators},
  author = {Souza, Ivo and Wilkens, Tim and Martin, Richard M.},
  year = 2000,
  month = jul,
  journal = {Phys. Rev. B},
  volume = {62},
  number = {3},
  pages = {1666--1683},
  issn = {0163-1829, 1095-3795},
  doi = {10.1103/PhysRevB.62.1666},
  urldate = {2024-01-22},
  langid = {english}
}

@article{Tang_EvidenceFrustratedMagnetic_2023,
  title = {Evidence of Frustrated Magnetic Interactions in a {{Wigner}}--{{Mott}} Insulator},
  author = {Tang, Yanhao and Su, Kaixiang and Li, Lizhong and Xu, Yang and Liu, Song and Watanabe, Kenji and Taniguchi, Takashi and Hone, James and Jian, Chao-Ming and Xu, Cenke and Mak, Kin Fai and Shan, Jie},
  year = 2023,
  month = mar,
  journal = {Nat. Nanotechnol.},
  volume = {18},
  number = {3},
  pages = {233--237},
  issn = {1748-3387, 1748-3395},
  doi = {10.1038/s41565-022-01309-8},
  urldate = {2024-03-19},
  langid = {english}
}

@article{Tsui_DirectObservationMagneticfieldinduced_2024,
  title = {Direct Observation of a Magnetic-Field-Induced {{Wigner}} Crystal},
  author = {Tsui, Yen-Chen and He, Minhao and Hu, Yuwen and Lake, Ethan and Wang, Taige and Watanabe, Kenji and Taniguchi, Takashi and Zaletel, Michael P. and Yazdani, Ali},
  year = 2024,
  month = apr,
  journal = {Nature},
  volume = {628},
  number = {8007},
  pages = {287--292},
  issn = {0028-0836, 1476-4687},
  doi = {10.1038/s41586-024-07212-7},
  urldate = {2025-12-11},
  langid = {english}
}

@article{Li_WignerMolecularCrystals_2024,
  title = {Wigner Molecular Crystals from Multielectron Moir\'e Artificial Atoms},
  author = {Li, Hongyuan and Xiang, Ziyu and Reddy, Aidan P. and Devakul, Trithep and Sailus, Renee and Banerjee, Rounak and Taniguchi, Takashi and Watanabe, Kenji and Tongay, Sefaattin and Zettl, Alex and Fu, Liang and Crommie, Michael F. and Wang, Feng},
  year = 2024,
  month = jul,
  journal = {Science},
  volume = {385},
  number = {6704},
  pages = {86--91},
  issn = {0036-8075, 1095-9203},
  doi = {10.1126/science.adk1348},
  urldate = {2024-09-18},
  langid = {english}
}

@article{Xiang_ImagingQuantumMelting_2025,
  title = {Imaging Quantum Melting in a Disordered {{2D Wigner}} Solid},
  author = {Xiang, Ziyu and Li, Hongyuan and Xiao, Jianghan and Naik, Mit H. and Ge, Zhehao and He, Zehao and Chen, Sudi and Nie, Jiahui and Li, Shiyu and Jiang, Yifan and Sailus, Renee and Banerjee, Rounak and Taniguchi, Takashi and Watanabe, Kenji and Tongay, Sefaattin and Louie, Steven G. and Crommie, Michael F. and Wang, Feng},
  year = 2025,
  month = may,
  journal = {Science},
  volume = {388},
  number = {6748},
  pages = {736--740},
  issn = {0036-8075, 1095-9203},
  doi = {10.1126/science.ado7136},
  urldate = {2026-02-05},
  langid = {english}
}

@misc{Ge_VisualizingImpactQuenched_2025,
  title = {Visualizing the {{Impact}} of {{Quenched Disorder}} on {{2D Electron Wigner Solids}}},
  author = {Ge, Zhehao and Smith, Conor and He, Zehao and Yang, Yubo and Li, Qize and Xiang, Ziyu and Xiao, Jianghan and Zhou, Wenjie and Kahn, Salman and Erdi, Melike and Banerjee, Rounak and Taniguchi, Takashi and Watanabe, Kenji and Tongay, Seth Ariel and Morales, Miguel A. and Zhang, Shiwei and Wang, Feng and Crommie, Michael F.},
  year = 2025,
  publisher = {arXiv},
  doi = {10.48550/ARXIV.2510.12009},
  urldate = {2026-01-16}
}

@misc{Berger_ImagingElectronHoleAsymmetry_2025,
  title = {Imaging {{Electron-Hole Asymmetry}} in the {{Quantum Melting}} of {{Generalized Wigner Crystals}}},
  author = {Berger, Emma and Arumainayagam, Michael and Dong, Zhihuan and Schneider, Lucas and Wang, Tianle and Nichols, Greyson and Kahn, Salman and Dutta, Rwik and Wang, Gaoqiang and Taniguchi, Takashi and Watanabe, Kenji and Naik, Mit H. and Zaletel, Michael P. and Wang, Feng and Crommie, Michael F.},
  year = 2025,
  publisher = {arXiv},
  doi = {10.48550/ARXIV.2512.16050},
  urldate = {2026-01-06}
}

@article{Goldshlager_2024,
   title={A Kaczmarz-inspired approach to accelerate the optimization of neural network wavefunctions},
   volume={516},
   ISSN={0021-9991},
   url={http://dx.doi.org/10.1016/j.jcp.2024.113351},
   DOI={10.1016/j.jcp.2024.113351},
   journal={Journal of Computational Physics},
   publisher={Elsevier BV},
   author={Goldshlager, Gil and Abrahamsen, Nilin and Lin, Lin},
   year={2024},
   month=nov, pages={113351} }

@article{Sorella_1998,
   title={Green Function Monte Carlo with Stochastic Reconfiguration},
   volume={80},
   ISSN={1079-7114},
   url={http://dx.doi.org/10.1103/PhysRevLett.80.4558},
   DOI={10.1103/physrevlett.80.4558},
   number={20},
   journal={Physical Review Letters},
   publisher={American Physical Society (APS)},
   author={Sorella, Sandro},
   year={1998},
   month=may, pages={4558–4561} }

@misc{chen2023efficientnumericalalgorithmlargescale,
      title={Efficient Numerical Algorithm for Large-Scale Damped Natural Gradient Descent}, 
      author={Yixiao Chen and Hao Xie and Han Wang},
      year={2023},
      eprint={2310.17556},
      archivePrefix={arXiv},
      primaryClass={cs.LG},
      url={https://arxiv.org/abs/2310.17556}, 
}

@article{Chen_2024,
   title={Empowering deep neural quantum states through efficient optimization},
   volume={20},
   ISSN={1745-2481},
   url={http://dx.doi.org/10.1038/s41567-024-02566-1},
   DOI={10.1038/s41567-024-02566-1},
   number={9},
   journal={Nature Physics},
   publisher={Springer Science and Business Media LLC},
   author={Chen, Ao and Heyl, Markus},
   year={2024},
   month=jul, pages={1476–1481} }

@ARTICLE{Zverevich2026,
       author = {{Zverevich}, Dmitry and {Levchenko}, Alex and {Esterlis}, Ilya},
        title = "{Spin-triplet paired Wigner crystal stabilized by quantum geometry}",
      journal = {arXiv e-prints},
         year = 2026,
        month = jan,
          eid = {arXiv:2601.05318},
        pages = {arXiv:2601.05318},
          doi = {10.48550/arXiv.2601.05318},
archivePrefix = {arXiv},
       eprint = {2601.05318},
 primaryClass = {cond-mat.str-el},
       adsurl = {https://ui.adsabs.harvard.edu/abs/2026arXiv260105318Z}
}

@article{CCK1976,
  title = {Exact calculations of the ground state of model neutron matter},
  author = {Ceperley, D. M. and Chester, G. V. and Kalos, M. H.},
  journal = {Phys. Rev. D},
  volume = {13},
  issue = {12},
  pages = {3208--3213},
  numpages = {0},
  year = {1976},
  month = {Jun},
  publisher = {American Physical Society},
  doi = {10.1103/PhysRevD.13.3208},
  url = {https://link.aps.org/doi/10.1103/PhysRevD.13.3208}
}

@article{besag1994comments,
  title={Comments on “Representations of knowledge in complex systems” by U. Grenander and MI Miller},
  author={Besag, Julian},
  journal={J. Roy. Statist. Soc. Ser. B},
  volume={56},
  number={591-592},
  pages={4},
  year={1994}
}

\end{document}